\documentclass[twocolumn,preprintnumbers,amsmath,amssymb,pra]{revtex4}

\usepackage{graphicx,amsmath}
\usepackage{dcolumn}
\usepackage{bm}
\usepackage{amssymb}
\usepackage{epstopdf}
\usepackage{color}
\usepackage[cp1251]{inputenc}
\usepackage[T2A]{fontenc}
\graphicspath{{fig_eps_new3/}}

\begin{document}

\title{Superradiant emission spectra of a two-qubit system in circuit quantum electrodynamics}

\begin{abstract}
In this paper we study the spontaneous emission spectra and the
emission decay rates of a simplest atom system that exhibits sub-
and superradiant properties: a system which consists of two
artificial atoms (superconducting qubits) embedded in a
one-dimensional open waveguide. The calculations are based on the
method of the transition operator which was firstly introduced by
R. H. Lehmberg  to theoretically describe the spontaneous emission
of two-level atoms in a free space. We obtain the explicit
expressions for the photon radiation spectra and the emission
decay rates for different initial two-qubit configurations with
one and two excitations. For every initial state  we calculate the
radiation spectra and the emission decay rates for different
effective distances between qubits. In every case, a decay rate is
compared with a single qubit decay to show the superradiant or
subradiant nature of a two-qubit decay with a given initial state.
\end{abstract}


\date{\today}

\author{Ya. S. Greenberg}\email{yakovgreenberg@yahoo.com}
\affiliation{Novosibirsk State Technical University, Novosibirsk,
Russia}
\author{O. A. Chuikin} \affiliation{Novosibirsk State
Technical University, Novosibirsk, Russia}


 \maketitle

\section{Introduction}
The control of spontaneous emission in the multi-atom (or qubit) system that interact
with a quantized radiation field in restricted geometries has received a great deal of
attention in recent years (see review paper \cite{Roy17} and references therein). This can
be achieved in various physical set-ups, for example, by putting two-level atoms in an
optical cavity \cite{Raimond01}, by embedding them in a nanophotonic waveguide
\cite{Noda07} or by coupling superconducting qubits to a transmission line resonator
\cite{Blais21, Wendin17, Gu17}. Due to spatial confinements, these set-ups allow one to
achieve an almost ideal mode matching which results in a strong coupling \cite{Wallraff04}
and even ultrastrong coupling regimes when the interaction strength overwhelms relaxation
rates \cite{Forn-Diaz19}. These experimental conditions are very challenging to obtain for
regular atoms in optical domain.

The interaction of excited atoms with the continuum of the
environment modes leads to spontaneous emission which is one of
the major sources of decoherence. The spontaneous emission results
in irreversible loss of the information encoded in the internal
states of the system and thus is regarded as the main obstacle in
practical implementations of quantum processing.

The early studies of spontaneous emission in multi- and two-atom
systems deal mainly with atoms placed in a free space environment
\cite{Lehmberg70a, Lehmberg70b, Ficek90, Rudolph95, Ficek02,
Lenz93}. In this case, the interaction of atoms with the vacuum
modes gives rise to short range dipole-dipole interaction between
the atoms \cite{Lehmberg70a}. The spontaneous emission of excited
atom in free space is the result of interaction of the atom with
the continuum modes of the vacuum. A collection of N identical two
level excited atoms undergoes a spontaneous coherent transition to
the ground state, which is accompanied by the emission of N
photons, the intensity of which scales as $N^2$, and the decay
rate of which is $N\Gamma$, where $\Gamma$ is the decay rate of an
isolated atom \cite{Dicke54}. Therefore, the N excited atoms decay
N times faster than an isolated atom. This property of N atom
system was named superradiance \cite{Gross82, Cong16}.

However, the behaviour of atoms in a confined geometry is quite
different from that in a free space. For example, a spontaneous
decay rate of an atom embedded in a resonator may significantly
differ from its decay rate in a free space (so called Purcell
effect) \cite{Purcell46}. The exchange of the virtual photons
between the identical equally spaced atoms in a one-dimensional
waveguide results in an infinite-range inter-atomic interaction
the strength of which periodically depends on the ratio
$d/\lambda$, where $d$ is the distance between neighbour atoms,
$\lambda$ is the wavelength of the guiding mode. Furthermore, this
system exhibits collective excitations with lifetimes from
extremely sub- to superradiant values relative to the radiative
lifetime of the individual atom \cite{Albrecht19, Zhang19,
Greenberg21, Brehm21, Mirhosseini19}.

Many of these effects have experimentally been realized within the
frame of circuit quantum electrodynamics (QED) with
superconducting qubits as artificial atoms \cite{Blais21}. A
significant difference between natural atoms and superconducting
qubits is that in circuit QED we can create artificial atoms with
desirable parameters some of which can be tuned, e.g. resonant
frequency or coupling strength \cite{You11}. What is more
important, we can address and manipulate the artificial atoms
individually \cite{Shevchenko19}. This makes it possible to
in-depth study of different types of interaction between a few
qubits which can be far more interesting and complex than just one
qubit in a cavity \cite{Ficek90, Ordonez04}. Because of the
perfect mode matching the exchange interaction between the qubits
is very strong and critically depends on the effective distance
between them which can be tuned by changing the wavelength via
qubit resonant frequency. Different spatial arrangements of qubits
in a chain can lead to significant modification of decay rates
\cite{Lalumiere13, vanLoo13}. This collective effect may lead to
both the enhancement of a decay rate, which is called
superradiance \cite{Delanty11, Lambert16}, and the reduction of
the decay rate, which corresponds to subradiance \cite{Albrecht19,
Zhang19}.

The arrangement consisting of two interacting atoms is a simplest
system which exhibits sub- and superradiant properties. This
system has been extensively studied in the frame of 1D circuit
quantum electrodynamics
\cite{Greenberg21,Mirhosseini19,Ordonez04,
Lalumiere13,vanLoo13,Delanty11,Kien2005,Mak2003}.

In this paper we consider a system which consists of two
artificial atoms (superconducting qubits) in a one-dimensional
open waveguide. In contrast to previous studies, we obtain here
the general explicit expression which allows us to calculate the
radiation photon spectra and the emission decay rates (decay rates
of the energy loss) for different initial two-qubit configurations
with one and two excitations and for different values of the ratio
$d/\lambda$. We systematically compare two-qubit spontaneous
emission spectra with those of a single qubit. We show that
depending on the ratio $d/\lambda$, there exist both superradiant
states when a decay rate of initial two-qubit configuration
exceeds that of a single qubit and subradiant states when a decay
rate of initial two-qubit configuration is less than that of a
single qubit. Our results for $d/\lambda\ll 1$ are consistent with
those which has experimentally been observed for two
superconducting qubits in a low quality cavity \cite{Mlynek14}.

There exist two approaches which allow us to perform these
calculations. The most common method uses the master equation for
the reduced density matrix in the Lindblad form
\cite{Lalumiere13}. The second approach uses the Heisenberg
equation of motion for arbitrary system operator
\cite{Lehmberg70a, Lehmberg70b}. Two approaches are, of course,
equivalent. The choice in favour of either of these methods
depends on the problem at hand.

Here we choose the second method and take as a system operator the
so-called transition operator, firstly introduced by Lehmberg
\cite{Lehmberg69} to theoretically describe the spontaneous
emission of two-level atoms in free space \cite{Lehmberg70a,
Lehmberg70b}. The exact expressions for the matrix elements of the
transition operator can be obtained within standard quantum
mechanics formalism using Heisenberg equations. The tracing out
the photonic modes from the equations of motion allows us to
obtain equations only for atomic operators independent of the
photon number. As distinct from conventional density matrix
approach, differential equations for the matrix elements of the
transition operator are linear in a basis set of a spin system
and, therefore, it is easier to solve them analytically. Moreover,
because the solutions of equations are operator functions, they
are independent of the specific initial state of the system. It
means that we don't need to find a new solution for every new
initial state like in the case of equations for the density
matrix. Once the exact solutions for transition operators are
found, we can use them not only to obtain transition probabilities
but also for the calculation of the photonic emission spectrum for
arbitrary initial density matrix. As we show here, transition
operators and elements of the density matrix are very closely
related, so one can easily switch from one approach to the other
one if needed.

The paper is organized as follows. In Sec. 2 we define the
transition operator, describe its general properties, and
establish its connection to a density matrix. In Sec. 3 we show
the application of the method for the description of N-qubit
system in a 1D transmission line and obtain general equation of
motion for the matrix elements of transition operators. In Sec. 4
we present a general expression for the spectral density of
photons in terms of two-time correlation functions of spin
operators and show how these types of correlation functions can be
calculated in terms of the vacuum average of transition operators
and initial density matrix. In Sec. 5 we apply our method to a
two-qubit system and find the diagonal and off-diagonal matrix
elements of the transition operator. In Sec. 6 we calculate the
probabilities of different transitions between the states in a
two-qubit system, which contribute to the radiation spectra and
the photon emission decay rates. The main results of our paper are
presented in Sec. 7. In this section we obtain the general
expressions which allow us to calculate both the radiation spectra
and the emission decay rates for arbitrary initial states of a
two-qubit system. For every initial state we consider in Sec. 7,
we calculate the radiation spectra and the emission decay rates
for different effective distances between qubits. In every case, a
decay rate is compared with a single qubit decay to show the
superradiant or subradiant nature of a two-qubit decay with a
given initial state. A summary of our work is presented in
concluding Sec. 8.

\section{General properties of transition operator}
We consider a system of $N$ identical qubits with eigenstates $\vert i\rangle$
coupled to a continuum of photon modes $\vert\nu\rangle$ in a one-dimensional
open waveguide. Our main interest is the probability of transition from some
arbitrary initial state of a qubit system $\vert\Psi_0\rangle$ with no photons
to some final state $\vert\Psi_1\rangle$ with $\nu$ photons in the field. According
to the general principles of quantum mechanics, the probability amplitude of such
 transition is given by the following matrix element:
\begin{equation}\label{2.1}
\left\langle {{\Psi _1},\nu } \right|{e^{ - iHt}}\left| {{\Psi _0},0} \right\rangle,
\end{equation}
where $H$ is the complete Hamiltonian which includes the N-qubit
system, photon field, and their interaction, $\vert0\rangle$,
$\vert\nu\rangle$ are the Fock states with zero and $\nu$ photons,
respectively. To find the total transition probability we must
find the squared modulus of this amplitude and sum it over the
complete set of possible final photon states $\vert\mu\rangle$ for
the field.

\begin{equation}\label{2.2}
W_{0 \to 1}  = \sum\limits_\mu  {\left| {\left\langle {\Psi _1
,\mu } \right|e^{ - iHt} \left| {\Psi _0 ,0} \right\rangle }
\right|^2 }
\end{equation}

 Using the completeness of the set of the photon states
$\sum\limits_\mu
{\left|\mu\right\rangle\left\langle\mu\right|}=1$, we rewrite
(\ref{2.2}) as follows:

\begin{equation}\label{2.2a}
W_{0 \to 1}  = \left\langle {\Psi _0 } \right|\left\langle 0
\right|e^{iHt} \left| {\Psi _1 } \right\rangle \left\langle {\Psi
_1 } \right|e^{ - iHt} \left| 0 \right\rangle \left| {\Psi _0 }
\right\rangle
\end{equation}

By expanding wave function $\vert\Psi_1\rangle$ over the complete
set of qubit states $\left|i\right\rangle$ with some coefficients
$c_i$  we can rewrite (\ref{2.2a}) in the following form:
\begin{equation}\label{2.4}
{W_{0 \to 1}} = \sum\limits_{i,j} {{c_i}c_j^ * } \left\langle {{\Psi _0}} \right|{\left\langle {{e^{iHt}}\left| i \right\rangle \left\langle j \right|{e^{ - iHt}}} \right\rangle _0}\left| {{\Psi _0}} \right\rangle,
\end{equation}
where $\langle\ldots\rangle_0$ is the average over photon vacuum
$\langle\ldots\rangle_0=\langle0\vert\ldots\vert0\rangle$

Following Lehmberg \cite{Lehmberg69}, we define a transition operator:
\begin{equation}\label{2.5}
{P_{i,j}}(t) = {e^{iHt}}\left|i\right\rangle\left\langle j\right|{e^{-iHt}}.
\end{equation}
The expression (\ref{2.4}) can then be written in a form:
\begin{equation}\label{2.6}
{W_{0\to1}} = \sum\limits_{i,j} {{c_i}c_j^*}\left\langle {{\Psi_0}}\right|{\left\langle {{P_{i,j}}(t)}\right\rangle_0}\left|{{\Psi_0}}\right\rangle.
\end{equation}
Thus, the probability of transition from the eigenstate $\vert n\rangle$ to the
eigenstate $\vert m\rangle$ can be calculated by using the corresponding matrix
element of the transition operator $P_{m,m}$ averaged over photon vacuum:
\begin{equation}\label{2.7}
{W_{n\to m}}=\left\langle n \right|{\left\langle{{P_{m,m}(t)}}\right\rangle_0}\left| n \right\rangle.
\end{equation}
It follows from the completeness of the qubits states $\vert i\rangle$ that the sum
of diagonal elements of the transition operator is equal to one:
\begin{equation}\label{2.8}
\sum\limits_i {{P_{i,i}} = 1}.
\end{equation}

As follows from the definition (\ref{2.5}), the expression for the elements of the
transition operator satisfy the Heisenberg equation:
\begin{equation}\label{2.9}
\frac{d}{{dt}}{P_{i,j}}(t) = i\left[ {H,{P_{i,j}}(t)} \right],
\end{equation}
with the initial conditions ${P_{i,j}}(0) = \left| i \right\rangle \left\langle j \right|$.

Unlike the equations for spin operators or elements of the density
matrix, equations for $P_{i,j}$ are linear: they contain only the
first degrees of the same operators. The number of equations for
the transition operator is determined by the number of states from
the complete set. For example, for one qubit there are only two
states, the excited state $|e\rangle$ and the ground state
$|g\rangle$. Accordingly, there are four matrix elements of the
transition operator: $P_{ee}$, $P_{gg}$, $P_{eg}$, and $P_{ge}$.
However, there are only three independent equations since the
$P_{eg}$ is a complex conjugate to a $P_{ge}$. For two qubits
there are four states and, consequently, there are ten independent
equations: four equations for the diagonal elements of the
transition operator and six equations for off-diagonal ones
(excluding complex conjugates).

In general case, the matrix elements of the transition operator
averaged over photon vacuum have the following form:
\begin{equation}\label{2.10}
{\left\langle {{P_{i,j}}(t)} \right\rangle _0} = \sum\nolimits_{m,n} {c_{mn}^{ij}} (t)\left| m \right\rangle \left\langle n \right|,
\end{equation}
where $c_{mn}^{ij}(t)$ are c-numbers.

In principle, we may choose the basis states of a spin system in
the following way:
\begin{equation}\label{2.10a}
\left\langle {P_{i,j} (t)} \right\rangle _0  = \sum\nolimits_{m,n}
{c_{mn}^{ij} } (t)\left| m \right\rangle \left\langle n
\right|;\quad i \ne j;\quad m \ne n
\end{equation}

\begin{equation}\label{2.10b}
\left\langle {P_{i,i} (t)} \right\rangle _0  = \sum\nolimits_q
{c_q^i } (t)\left| q \right\rangle \left\langle q \right|
\end{equation}

In this case, the off-diagonal element,  Eq. \ref{2.10a}, of
transition operator provides the transitions between different
states of a spin system, while the diagonal element of transition
operator has only diagonal matrix elements in the Hilbert space of
a spin system.

By the definition, the qubit density matrix of a spin system,
$\rho_{S,ji}^{}(t) = \left\langle j\right|T{r_\nu}\left[ {\rho(t)}
\right]\left|i\right\rangle \equiv \left\langle
j\right|{\rho_S}(t)\left| i \right\rangle$, where $\rho(t)$ is the
density matrix of the whole system:
\begin{equation}\label{2.11}
\rho (t) = {e^{ - iHt}}\rho \,(0){e^{iHt}},
\end{equation}
can be expressed in terms of the average value of the transition operator:
\begin{equation}\label{2.12}
\rho _{S,ji}^{}(t) = T{r_{S,\nu }}\left( {\rho (0){P_{i,j}}(t)} \right) \equiv \left\langle {{P_{i,j}}(t)} \right\rangle,
\end{equation}
where the trace is taken over both the qubit system and the photon
field, and $\rho(0)$ is the initial density matrix of the whole
system.

If we assume that initially the field is in a photon vacuum:
\begin{equation}\label{2.13}
\rho (0) = {\rho _S}(0) \otimes {\rho _\nu }(0) = {\rho _S}(0)
\otimes \left| 0 \right\rangle \left\langle 0 \right|,
\end{equation}
we then obtain from (\ref{2.12}):
\begin{equation}\label{2.14}
\left\langle l \right|{\rho _S}(t)\left| m \right\rangle  = \sum\limits_{n.,q} {\left\langle n \right|{\rho _S}(0)\left| q \right\rangle \left\langle q \right|\langle{P_{m,l}}(t)\rangle_0 \left| n \right\rangle }.
\end{equation}

Therefore, the matrix elements of a reduced density matrix can be
expressed in terms of  the matrix elements of the vacuum average
of the transition operator if the initial density matrix
$\rho_S(0)$ is known. It should be noted that in contrast to the
elements of the density matrix, which are numerical functions, the
matrix elements of $P_{i,j}(t)$ are the operator functions. If the
system is initially in one of its basis states
$|s\rangle$($\langle|s|\rho_S(0)|s\rangle=1$), then it follows
from (\ref{2.13}):
\begin{equation}\label{2.13a}
\left\langle l \right|\rho _S (t)\left| m \right\rangle  =
\left\langle s \right|\left\langle {P_{m,l} (t)} \right\rangle _0
\left| s \right\rangle
\end{equation}
For basis set with the properties (\ref{2.10a}) and (\ref{2.10b})
it follows from (\ref{2.13a}) that the off-diagonal elements of
reduced density matrix are zero. For the diagonal elements of
reduced density matrix, that is, for the populations, we obtain:.
\begin{equation}\label{2.13b}
\left\langle m \right|\rho _S (t)\left| m \right\rangle  =
\left\langle s \right|\left\langle {P_{m,m} (t)} \right\rangle _0
\left| s \right\rangle
\end{equation}
Therefore, if $m\neq s$ the population $\left\langle m \right|\rho
_S (t)\left| m \right\rangle$   can be understood as the
transition amplitude from the initial state $ |s\rangle$  to the
state $|m\rangle$.


\section{Multi-qubit system}

Consider a system consisting of N qubits in a one-dimensional infinite waveguide. This
system can be described by a Jaynes-Cummings Hamiltonian:
\begin{multline}\label{3.1}
H = \frac{1}{2}\sum\limits_{n = 1}^N {\left( {1 + \sigma _z^{(n)}} \right){\Omega _n}}  + \sum\limits_k {{\omega _k}a_k^\dag {a_k}} \\
 + \sum\limits_k {\left( {a_k^\dag S_k^ -  + S_k^ + {a_k}} \right)},
\end{multline}
where we introduced collective atomic spin operators:
\begin{equation}\label{3.2}
S_k^ -  = \sum\limits_{n = 1}^N {g_k^{(n)}} {e^{ - ik{x_n}}}\sigma _ - ^{(n)},\quad \quad S_k^ +  = \sum\limits_{n = 1}^N {g_k^{*(n)}} {e^{ik{x_n}}}\sigma _ + ^{(n)},
\end{equation}
Here $\sigma^{(n)}_z$ is a Pauli spin operator, $\Omega_n$ is a
resonant frequency of nth qubit, $a_k^\dag$($a_k$) are creation
(annihilation) operators for a $k$ mode photon, $\omega_k$ is a
photon frequency, $\sigma_-^{(n)} = {\left|g\right\rangle_{nn}}
\left\langle e \right|$ and $\sigma _ + ^{(n)} = {\left| e
\right\rangle _{nn}} \left\langle g \right|$ are the atomic ladder
operators which lower or raise a state of the nth qubit, $g_k$ is
a coupling strength between the qubit and the field, $x_n$ is a
spatial coordinate of the nth qubit.

From (\ref{2.9}) and (\ref{3.1}) we obtain the equation of motion for the transition operator:
\begin{equation}\label{3.3}
\begin{gathered}
\frac{{d}}{{dt}}{P_{ij}} = \\
\frac{i}{2}\sum\limits_{n = 1}^N {{\Omega _n}} \left( {{e^{iHt}}\sigma _Z^{(n)}\left| i \right\rangle \left\langle j \right|{e^{ - iHt}} - {e^{iHt}}\left| i \right\rangle \left\langle j \right|\sigma _Z^{(n)}{e^{ - iHt}}} \right)\\
 + i\sum\limits_k {a_k^\dag } (t)\left( {{e^{iHt}}S_k^ - \left| i \right\rangle \left\langle j \right|{e^{ - iHt}} - {e^{iHt}}\left| i \right\rangle \left\langle j \right|S_k^ - {e^{ - iHt}}} \right)\\
 + i\sum\limits_k {\left( {{e^{iHt}}S_k^ + \left| i \right\rangle \left\langle j \right|{e^{ - iHt}} - {e^{iHt}}\left| i \right\rangle \left\langle j \right|S_k^ + {e^{ - iHt}}} \right)} \,{a_k}(t),
\end{gathered}
\end{equation}
where the photon operators are in the Heisenberg representation:
\begin{equation}\label{3.4}
a_k^\dag (t) = {e^{iHt}}a_k^\dag {e^{ - iHt}},\;\quad {a_k}(t) = {e^{iHt}}{a_k}{e^{ - iHt}}.
\end{equation}

For photon operators the equations of motion are as follows:
\begin{subequations} \label{3.5}
\begin{align}
i\frac{{d{a_k}}}{{dt}} &= \left[ {{a_k}(t),H} \right] = {\omega _k}{a_k}(t) + S_k^ - (t),\label{3.5a}
\\
i\frac{{da_k^\dag }}{{dt}} &= \left[ {a_k^\dag (t),H} \right] =  - {\omega _k}a_k^\dag (t) - S_k^ + (t).\label{3.5b}
\end{align}
\end{subequations}
where $S_k^ \pm (t)$ are collective spin operators in the Heisenberg picture. The formal
solution of these equations is given by:
\begin{subequations}\label{3.6}
\begin{align}
{a_k}(t) &= {a_k}(0){e^{ - i{\omega _k}t}} - i\int\limits_0^t {{e^{ - i{\omega _k}(t - \tau )}}S_k^ - (\tau )d\tau },\label{3.6a}
\\
a_k^\dag (t) &= a_k^\dag (0){e^{i{\omega _k}t}} + i\int\limits_0^t {{e^{i{\omega _k}(t - \tau )}}S_k^ + (\tau )d\tau }.\label{3.6b}
\end{align}
\end{subequations}
where the first term in the right hand side of (\ref{3.6}) is a
free field part and the second term is a part radiated by atoms.

Since we considering a 1D waveguide, $k$ takes only two directions, $k=\pm
|k|=\pm\omega/v_g$. From (\ref{3.2}) it follows that positive $k$,
$|k|=+\omega/v_g$, corresponds to the right (forward)  propagating
modes, while the negative  $k$, $|k|=-\omega/v_g$, corresponds to
the  left (backward) propagating modes. Therefore, the operators
$a_{+|k|}, a^+_{+|k|}$, and $a_{-|k|}, a^+_{-|k|}$ correspond to
right and left   propagating photons, respectively.

 Even though the photon operators $a_k^\dag (t),\;\,{a_k}(t)$
commute with collective spin operators $S_k^ \pm (t)$, each term
in (\ref{3.6}) does not commute with $S_k^ \pm (t)$. This explains
the position of these operators in the third term of Hamiltonian
(\ref{3.1}). They should be placed in such a way that in the final
expression the creation operators $a_k^\dag (t)$ were placed on
the left of the transition operator $P_{i,j}$, while the
annihilation operators ${a_k}(t)$ were placed on the right. This
is necessary for the terms with initial photons to be dropped out
upon averaging the transition operator over the photon vacuum.

Substituting the expressions (\ref{3.6}) into the equation of
motion (\ref{3.3}) we obtain:
\begin{widetext}
\begin{equation}\label{3.7}
\begin{gathered}
\frac{{d{P_{ij}}}}{{dt}} = i\frac{1}{2}\sum\limits_{n = 1}^N {{\Omega _n}} \left( {{e^{iHt}}\sigma _Z^{(n)}\left| i \right\rangle \left\langle j \right|{e^{ - iHt}} - {e^{iHt}}\left| i \right\rangle \left\langle j \right|\sigma _Z^{(n)}{e^{ - iHt}}} \right)\\
 + i\sum\limits_k {a_k^\dag (0)} {e^{i{\omega _k}t}}{e^{iHt}}S_k^ - \left| i \right\rangle \left\langle j \right|{e^{ - iHt}} - i\sum\limits_k {a_k^\dag (0)} {e^{i{\omega _k}t}}{e^{iHt}}\left| i \right\rangle \left\langle j \right|S_k^ - {e^{ - iHt}}\\
 + i\sum\limits_k {{e^{iHt}}} S_k^ + \left| i \right\rangle \left\langle j \right|{e^{ - iHt}}{a_k}(0){e^{ - i{\omega _k}t}} - i\sum\limits_k {{e^{iHt}}} \left| i \right\rangle \left\langle j \right|S_k^ + {e^{ - iHt}}{a_k}(0){e^{ - i{\omega _k}t}}\\
 + \sum\limits_k {\int\limits_0^t {{e^{i{\omega _k}(t - \tau )}}S_k^ + (\tau )d\tau \;} } {e^{iHt}}\left[ {\left| i \right\rangle \left\langle j \right|,S_k^ - } \right]{e^{ - iHt}} + \sum\limits_k {{e^{iHt}}} \left[ {S_k^ + ,\left| i \right\rangle \left\langle j \right|} \right]{e^{ - iHt}}\int\limits_0^t {{e^{ - i{\omega _k}(t - \tau )}}S_k^ - (\tau )d\tau }
 \end{gathered}
\end{equation}
\end{widetext}
Here we write the equation in such a manner that the action of atomic operators on system
states is clearly visible. Equation (\ref{3.7}) can be rewritten in terms of transition
operators and atomic operators in the Heisenberg picture, since ${e^{iHt}}\sigma_Z^{(n)}
\left|i\right\rangle\left\langle j\right|{e^{-iHt}}=\sigma_Z^{(n)}(t){P_{i,j}}(t)$ (the
same procedure applies to the terms with $S_k^\pm$ as well).

Up to now, we did not make any approximations: the above expressions are exact. In order
to solve the equation (\ref{3.7}), the following assumptions are made:
\begin{subequations}\label{3.8}
\begin{align}
\sigma _ - ^{(n)}(\tau ) &\approx \sigma _ - ^{(n)}(t){e^{ - i{\Omega _n}(\tau  - t)}},\label{3.8a}
\\
\sigma _ + ^{(n)}(\tau ) &\approx \sigma _ + ^{(n)}(t){e^{i{\Omega _n}(\tau  - t)}}.\label{3.8b}
\end{align}
\end{subequations}
Assuming that all qubit frequencies are identical and equal to
some value $\Omega$, we then obtain:
\begin{subequations}\label{3.9}
\begin{align}
S_k^ - (\tau ) &\approx S_k^ - (t){e^{ - i\Omega (\tau  - t)}},\label{3.9a}
\\
S_k^ + (\tau ) &\approx S_k^ + (t){e^{i\Omega (\tau  - t)}}.\label{3.9b}
\end{align}
\end{subequations}

The assumptions (\ref{3.8}), (\ref{3.9}) are equivalent to
Wigner-Weisskopf or Markov approximations
\cite{Lehmberg70a,Lalumiere13}. It allows us to take $S_k^ \pm
(\tau )$ out of the integrand in the last line of (\ref{3.7}). We
then rewrite the rest of the integrals by taking into account the resonant
approximation, i.e. assuming that the main contribution to the
integral is near the resonance frequency $\Omega$. This allows us
to take the upper limit to infinity, and we get:
\begin{multline}\label{3.10}
\int\limits_0^t {{e^{i(\omega  - \Omega )(t - t')}}dt'}  \approx \int\limits_0^\infty  {{e^{i(\omega  - \Omega )\tau }}d\tau }  = \\
\pi \delta (\omega  - \Omega )+ i\,P.v.\left( {\frac{1}{{\omega  - \Omega }}} \right),
\end{multline}
where $\delta(\omega)$ is a Dirac delta function and $P.v.$ is a Cauchy principal value.

According to all assumptions above, we obtain from (\ref{3.7}) the equation of motion for
the transition operator in the following form:
\begin{widetext}
\begin{equation}\label{3.11}
\begin{gathered}
\frac{{d{P_{ij}}}}{{dt}} = i\frac{1}{2}\sum\limits_{n = 1}^N {{\Omega _n}} \left( {{e^{iHt}}\sigma _Z^{(n)}\left| i \right\rangle \left\langle j \right|{e^{ - iHt}} - {e^{iHt}}\left| i \right\rangle \left\langle j \right|\sigma _Z^{(n)}{e^{ - iHt}}} \right)\\
 + i\sum\limits_k {a_k^\dag } (0){e^{i{\omega _k}t}}{e^{iHt}}\left[ {S_k^ - ,\left| i \right\rangle \left\langle j \right|} \right]{e^{ - iHt}} + i\sum\limits_k {{e^{iHt}}} \left[ {S_k^ + ,\left| i \right\rangle \left\langle j \right|} \right]{e^{ - iHt}}{a_k}(0){e^{ - i{\omega _k}t}}\\
 + \sum\limits_{n,m}^N {\frac{{{\Gamma _{n,m}}}}{2}} \,{e^{iHt}}\left( {2\sigma _ + ^{(m)}\left| i \right\rangle \left\langle j \right|\sigma _ - ^{(n)} - \sigma _ + ^{(m)}\sigma _ - ^{(n)}\left| i \right\rangle \left\langle j \right| - \left| i \right\rangle \left\langle j \right|\sigma _ + ^{(m)}\sigma _ - ^{(n)}} \right){e^{ - iHt}}\\
 + i\sum\limits_{n,m}^N {{\alpha _{n,m}}} \,{e^{iHt}}\left( {\left| i \right\rangle \left\langle j \right|\sigma _ + ^{(m)}\sigma _ - ^{(n)} - \sigma _ + ^{(m)}\sigma _ - ^{(n)}\left| i \right\rangle \left\langle j \right|} \right){e^{ - iHt}}
\end{gathered}
\end{equation}
\end{widetext}
where according to the Fermi Golden rule we have introduced a decay rate $\Gamma_{n,m}$:
\begin{equation}\label{3.12a}
{\Gamma _{n,m}} = \sum\limits_k {2\pi } \delta \left( {{\omega _k} - \Omega } \right)g_k^{(n)}g_k^{*(m)}{e^{ - ik({x_n} - {x_m})}},
\end{equation}
and a frequency shift $\alpha_{n,m}$:
\begin{equation}\label{3.12b}
{\alpha _{n,m}} = \sum\limits_k {g_k^{(n)}} g_k^{*(m)}{e^{ - ik({x_n} - {x_m})}}P.v.\left( {\frac{1}{{{\omega _k} - \Omega }}} \right).
\end{equation}
Equation (\ref{3.11}) is the most general case of the equation of
motion for the matrix elements of the transition operator
(\ref{2.5}) for $N$ qubits with identical resonant frequencies.
When equation (\ref{3.11}) is averaged over the initial photon
vacuum, the second line in (\ref{3.11}) can be dropped out.

For a long 1D waveguide we can replace the summation over $k$ by the integration:
\begin{equation}\label{3.13}
\sum\limits_k {f(k)}  \to \frac{L}{{2\pi }}\int\limits_{ - \infty
}^{ + \infty }{f(k)} {d\left| k \right|}  = \frac{L}{{\pi
{\upsilon _g}}}\int\limits_0^\infty {(f(|k|)+f(-|k|))} {d\omega }
\end{equation}
where $L$ is the quantization length in the propagation direction and $\upsilon_g$ is
the photon group velocity.

If we assume that the coupling strength is the same for all qubits, $g_k^{(n)}=g_k^{(m)}
\equiv{g_k}$ and is symmetrical, i.e. $g_k = g_{-k}$, and it contributes mainly near the
resonance $g_k\approx g_{k_0}$, where $k_0=\Omega/\upsilon_g$, we then obtain for
(\ref{3.12a}) and (\ref{3.12b}):
\begin{subequations}\label{3.14}
\begin{align}
{\Gamma _{n,m}} &= \Gamma \cos ({k_0}{\kern 1pt} \left| {{d_{n,m}}} \right|),\label{3.14a}
\\
{\alpha _{n,m}} &=  - \frac{\Gamma }{2}\sin ({k_0}{\kern 1pt} \left| {{d_{n,m}}} \right|),\label{3.14b}
\end{align}
\end{subequations}
where $d_{n,m} = (x_n - x_m)$ is the distance between nth and mth
qubit, $\Gamma$ is the single-qubit emission rate into the
waveguide mode:

\begin{equation}\label{3.14c}
\Gamma  = \frac{{2L}}{{{\upsilon _g}}}{\left| {g_{{k_0}}^{}} \right|^2},
\end{equation}
The expression (\ref{3.14a}) is obtained with the help of the
following relation:
\begin{equation}\label{3.15}
P.v.\int\limits_0^{+\infty} {\frac{\cos({{\omega{d_{n,m}}}/{\upsilon _g}})}{\omega -\Omega }} \,d\omega  = -\pi \sin \left( {{k_0}\left|{{d_{n,m}}}\right|}\right).
\end{equation}
The expression (\ref{3.15}) is exact if  counter-rotating terms in
the qubit-field interaction is taken into account (Suppl. in
\cite{Gonz2013}). Nevertheless, within a rotating wave
approximation the Eq. \ref{3.15} provides a good accuracy for
$d>\lambda/4$ \cite{Greenberg21}.

The quantities $\Gamma_{n,m}$ and $\alpha_{n,m}$ denote the
dissipative and coherent interaction rates, respectively. The
coherent interaction results from the exchange of virtual photons
between qubits at all continuum frequencies except for a single
frequency $\omega=\Omega$. It gives rise to the shift of the qubit
frequencies. In contrast to the case of a free space, these
inter-qubit interactions have an infinite range. In addition, the
interaction between any two qubits can be easily switched off by a
proper choice of the value $k_0|d_{n,m}|$. In the simple case for
which a distance between any two neighbor qubits is $d$, the
coherent inter qubit interaction vanish if $k_0d=n\pi$ where $n$
is any positive integer.

\section{Radiation spectrum and calculation of the correlation functions}

In circuit quantum electrodynamics it is possible to
experimentally measure both the full photon spectrum $\left\langle
{a_k^\dag(t){a_k}(t)}\right\rangle$ and one-time mean values of
single-photon operators
$\left\langle{{a_k}(t)}\right\rangle,\;\left\langle
{a_k^+(t)}\right\rangle$ \cite{Caves1982,Eich2012}. Moreover, we
can construct more complex photon correlation functions with a
higher order of photon operators and experimentally measure them
as well \cite{Kannan20}. In this paper, the main attention is paid
to the quantity $\left\langle {a_k^\dag(t){a_k}(t)}\right\rangle$
which defines the photon radiation spectrum.

From the expressions for photon operators (\ref{3.6}) we obtain:

\begin{equation}\label{4.2}
\left\langle {a_k^\dag (t){a_k}(t)} \right\rangle
 = \int\limits_0^t {d\tau } \int\limits_0^t {d\tau '}
  \;{e^{ - i{\omega _k}(\tau  - \tau ')}}\left\langle {S_k^ + (\tau )S_k^ - (\tau ')} \right\rangle.
\end{equation}

From (\ref{4.2}) we obtain the total photon emission rate, that
is, the rate of the energy loss:
\begin{equation}\label{4.2a}
W(t) = \frac{d} {{dt}}\sum\limits_k {\left\langle {a_k^\dag (t)a_k
(t)} \right\rangle }  = \left\langle {S_k^ +  (t)S_k^ - (t)}
\right\rangle
\end{equation}

The frequency dependent photon radiation spectrum is defined as
the limit of (\ref{4.2}) when $t$ tends to infinity:

\begin{equation}\label{4.2d}
S(\omega_k)
 = \int\limits_0^\infty {d\tau } \int\limits_0^\infty {d\tau '}
  \;{e^{ - i{\omega _k}(\tau  - \tau ')}}\left\langle {S_k^ + (\tau )S_k^ - (\tau ')}
   \right\rangle.
\end{equation}

The averaging in above equations is understood as the tracing over
both the states $S$ of a spin system  and the states $\nu$ of a
photon field.

\begin{equation}\label{trace}
    \left\langle {S_k^ + (\tau )S_k^ - (\tau ')}
   \right\rangle=Tr_{S,\nu}\left({S_k^ + (\tau )S_k^ - (\tau
   ')\rho(0)}\right)
\end{equation}
where $\rho(0)$ is the initial density matrix of the whole system.

From definition (\ref{3.2}) we obtain:
\begin{multline}\label{4.8}
\left\langle {S_k^ + (\tau )S_k^ - (\tau ')} \right\rangle  = \\
\sum\limits_{n,m}^N {g_k^{(m)}g_k^{*(n)}{e^{ik({x_n} - {x_m})}}} \left\langle {\sigma _ + ^{(n)}(\tau )\sigma _ - ^{(m)}(\tau ')} \right\rangle.
\end{multline}

The expression for two-time correlation function $\left\langle {\sigma _ + ^{(n)}(\tau )\sigma _ - ^{(m)}(\tau ')} \right\rangle $ can be written in two ways depending on the relation between $\tau$ and $\tau^\prime$. If $\tau>\tau^\prime$, then:
\begin{subequations}\label{4.9}
\begin{align}
\left\langle {\sigma _ + ^{(n)}(\tau )\sigma _ - ^{(m)}(\tau ')} \right\rangle = T{r_{S,\nu }}\left[ {\rho (\tau ')\sigma _ + ^{(n)}(\tau  - \tau ')\sigma _ - ^{(m)}(0)} \right].\label{4.9a}
\\
\intertext{If $\tau<\tau^\prime$, then:}
\left\langle {\sigma _ + ^{(n)}(\tau )\sigma _ - ^{(m)}(\tau ')} \right\rangle  = T{r_{S,\nu }}\left[ {\sigma _ + ^{(m)}(0)\sigma _ - ^{(n)}(\tau ' - \tau )\rho (\tau )} \right].\label{4.9b}
\end{align}
\end{subequations}
These prescriptions come from the requirement for the time argument of a spin operator
to be always positive.

The expressions (\ref{4.9}) are exact. If we assume the system is
always in a photon vacuum state (\ref{2.13}) we obtain for
(\ref{4.9a}), (\ref{4.9b}):


\begin{subequations}\label{4.11}
\begin{align}
&\begin{gathered}
\left\langle {\sigma _ + ^{(n)}(\tau )\sigma _ - ^{(m)}(\tau ')} \right\rangle  = \\
T{r_S}\left[ {{\rho _S}(\tau '){{\left\langle {\sigma _ + ^{(n)}(\tau  - \tau ')} \right\rangle }_0}\sigma _ - ^{(m)}(0)} \right],
\end{gathered} \quad \text{$\tau>\tau'$} \label{4.11a}
\\
&\begin{gathered}
\left\langle {\sigma _ + ^{(n)}(\tau )\sigma _ - ^{(m)}(\tau ')} \right\rangle  = \\
T{r_S}\left[ {\sigma _ + ^{(m)}(0){{\left\langle {\sigma _ - ^{(n)}(\tau ' - \tau )} \right\rangle }_0}{\rho _S}(\tau )} \right],
\end{gathered} \quad \quad \text{$\tau<\tau'$}\label{4.11b}
\end{align}
\end{subequations}

\begin{widetext}
With the aid of (\ref{2.14}) we can rewrite the density matrix
$\rho_S(t)$ in (\ref{4.11}) in terms of the initial density matrix
$\rho_S(0)$ and transition operators:

\begin{subequations}\label{4.13}
\begin{align}
\begin{gathered}
\left\langle {\sigma _+^{(n)}(\tau)\sigma_-^{(m)}(\tau')}
\right\rangle_+
 = \sum\limits_{l,q} {\sum\limits_{s,p} {} \left\langle s \right|{\rho _S}(0)\left| p \right\rangle \left\langle p \right|{{\left\langle {{P_{q,l}}(\tau ')} \right\rangle }_0}\left| s \right\rangle \left\langle q \right|{{\left\langle {\sigma _ + ^{(n)}(\tau  - \tau ')} \right\rangle }_0}\sigma _ - ^{(m)}(0)\left| l \right\rangle },
\end{gathered} \quad\quad \text{$\tau>\tau'$}\label{4.13a}
\\
\begin{gathered}
\left\langle {\sigma _+^{(n)}(\tau)\sigma_-^{(m)}(\tau')}
\right\rangle_- = \sum\limits_{l,q} {\sum\limits_{s,p} {}
\left\langle l \right|\sigma _ + ^{(n)}(0){{\left\langle {\sigma _
- ^{(m)}(\tau ' - \tau )} \right\rangle }_0}\left| q \right\rangle
\left\langle s \right|{\rho _S}(0)\left| p \right\rangle
\left\langle p \right|{{\left\langle {{P_{lq}}(\tau )}
\right\rangle }_0}\left| s \right\rangle }.
\end{gathered} \quad\quad \text{$\tau<\tau'$} \label{4.13b}
\end{align}
\end{subequations}
Here plus and minus subscript just indicate the positive and
negative time difference. The quantity ${\left\langle {\sigma _
\pm ^{(n)}(\tau  - \tau ')} \right\rangle _0}$ in (\ref{4.13}),
which refers to the individual nth spin of a system should be
expressed in terms of the matrix elements of transition operator
acting in the collective basis set of a spin system.

Putting all things together, we have the following expression for
spectrum (\ref{4.2}):
\begin{equation}\label{4.12}
\begin{gathered}
\left\langle {a_k^\dag (t){a_k}(t)} \right\rangle  =
\sum\limits_{n,m}^N {g_k^{(m)}g_k^{*(n)}{e^{ik({x_n} - {x_m})}}}
\int\limits_0^t {d\tau } \int\limits_0^\tau  {d\tau '} {e^{ -
i{\omega _k}(\tau  - \tau ')}} \left\langle {\sigma
_+^{(n)}(\tau)\sigma_-^{(m)}(\tau')} \right\rangle_+
\\
 + \sum\limits_{n,m}^N {g_k^{(m)}g_k^{*(n)}{e^{ik({x_n} - {x_m})}}} \int\limits_0^t {d\tau } \int\limits_\tau ^t {d\tau '} {e^{ - i{\omega _k}(\tau  - \tau ')}}
 \left\langle {\sigma _+^{(n)}(\tau)\sigma_-^{(m)}(\tau')} \right\rangle_-
\end{gathered}
\end{equation}

where two-time correlation functions are given in (\ref{4.13}) for
different time intervals. Therefore, we can calculate the desired
correlation functions using only transition operators averaged
over the vacuum field.
\end{widetext}

Note that this approach allows one to calculate not only two-time
correlation functions similar to (\ref{4.9}) but more complex ones
as well. In (\ref{4.13}) we reduce the averaged value of two
operators to an average of one operator multiplied by the second
operator at zero time. The same principle can be used to reduce,
say, a four-time correlation function to a three-time correlation
function, which can be also reduced the same way to a two-time
correlation function, and so on. Thus, we can find higher-order
correlation functions using only transition operators found from
(\ref{3.11}).

\section{Transition operators for two-qubit system}

For two-qubit system there are four basis states:
\begin{equation}\label{6.1}
\left| 1 \right\rangle  = \left| {gg} \right\rangle ;\quad \;\left| 2 \right\rangle  = \left| {ee} \right\rangle ;\;\quad \left| 3 \right\rangle  = \left| {ge} \right\rangle ;\;\quad \left| 4 \right\rangle  = \left| {eg} \right\rangle
\end{equation}
However, we use here a so called Dicke basis consisting of states
$\vert 1\rangle$, $\vert 2\rangle$ and symmetrical and
asymmetrical superposition of states $\vert 3\rangle$ and $\vert
4\rangle$:
\begin{equation}\label{6.2}
\begin{gathered}
\left| G \right\rangle  = \left| {gg} \right\rangle, \quad \left| E \right\rangle  = \left| {ee} \right\rangle, \\
\left| S \right\rangle  = \frac{1}{{\sqrt 2 }}\left( {\left| {ge} \right\rangle  + \left| {eg} \right\rangle } \right), \quad \quad \left| A \right\rangle  = \frac{1}{{\sqrt 2 }}\left( {\left| {ge} \right\rangle  - \left| {eg} \right\rangle } \right).
\end{gathered}
\end{equation}
The advantage of basis states (\ref{6.2}) over the (\ref{6.1}) is that the equations of
motion for diagonal matrix elements of transition operator are independent of the
off-diagonal ones.

Using the definition of lowering and raising operators for the regular basis (\ref{6.1}),
it is easy to show how they act on the new basis states (\ref{6.2}):
\begin{subequations}\label{6.3}
\begin{align}
\begin{gathered}
\sigma _ + ^{(1,2)}\left| G \right\rangle  = \frac{1}{{\sqrt 2 }}\left( {\left| S \right\rangle  \mp \left| A \right\rangle } \right),\quad \quad \sigma _ - ^{(1,2)}\left| G \right\rangle  = 0,\\
\sigma _ + ^{(1,2)}\left| E \right\rangle  = 0,\quad \quad \sigma _ - ^{(1,2)}\left| E \right\rangle  = \frac{1}{{\sqrt 2 }}\left( {\left| S \right\rangle  \pm \left| A \right\rangle } \right),
\end{gathered}\label{6.3a} \\
\begin{gathered}
\sigma _ + ^{(1,2)}\left| S \right\rangle  = \frac{1}{{\sqrt 2 }}\left| E \right\rangle ,\quad \quad \sigma _ - ^{(1,2)}\left| S \right\rangle  = \frac{1}{{\sqrt 2 }}\left| G \right\rangle, \\
\sigma _ + ^{(1,2)}\left| A \right\rangle  =  \pm \frac{1}{{\sqrt 2 }}\left| E \right\rangle ,\quad \quad \sigma _ - ^{(1,2)}\left| A \right\rangle  =  \mp \frac{1}{{\sqrt 2 }}\left| G \right\rangle.
\end{gathered}\label{6.3b}
\end{align}
\end{subequations}
The same can be easily done for Pauli spin operators:
\begin{equation}\label{6.4}
\begin{gathered}
\sigma _Z^{(1,\,2)}\left| G \right\rangle  =  - \left| G \right\rangle ,\quad \sigma _Z^{(1,\,2)}\left| E \right\rangle  = \left| E \right\rangle ,\\
\sigma _Z^{(1,\,2)}\left| A \right\rangle  =  \mp \left| S \right\rangle ,\quad \sigma _Z^{(1,\,2)}\left| S \right\rangle  =  \mp \left| A \right\rangle .
\end{gathered}
\end{equation}
Next, we apply the equation (\ref{3.11}) for $N = 2$ and assume
for the decay rates, ${\Gamma_{11}}={\Gamma_{22}}= \Gamma$,
${\Gamma_{12}}={\Gamma_{21}}=\Gamma\cos({k_0}d)$, and for the
frequency shifts ${\alpha_{11}}={\alpha_{22}} = 0$, ${\alpha_{12}}
= {\alpha_{21}} = {\Gamma\sin({k_0}d)} /2$, where $d$ is the
distance between two qubits. By averaging the equation
(\ref{3.11}) over the photon vacuum state $\vert 0\rangle$, the
terms including photon operators in the second line in
(\ref{3.11}) will be dropped out, and we can obtain equations for
the matrix elements of the transition operator. For the basis
(\ref{6.2}) we have sixteen equations in total, but since the
off-diagonal transition operators $P_{i,j}$ is a hermitian
conjugate of $P_{j,i}$, it is sufficient to find the solution only
for ten matrix elements of the transition operator. For the
diagonal matrix elements of transition operator (which we will
refer to as populations by analogy with diagonal elements of
density matrix), we find:
\begin{subequations}\label{6.6}
\begin{align}
\frac{{d{{\left\langle {{P_{EE}}} \right\rangle }_0}}}{{dt}} &=  - 2\Gamma {\left\langle {{P_{EE}}} \right\rangle _0},\label{6.6a}
\\
\frac{{d{{\left\langle {{P_{SS}}} \right\rangle }_0}}}{{dt}} &= \Gamma \left({1+\cos{k_0}d}\right)\left({{\left\langle {{P_{EE}}} \right\rangle _0} + {\left\langle {P_{SS}} \right\rangle_0} }\right),\label{6.6b}
\\
\frac{{d{{\left\langle {{P_{AA}}} \right\rangle }_0}}}{{dt}} &= \Gamma \left({1-\cos{k_0}d}\right)\left({{\left\langle {{P_{EE}}} \right\rangle _0} - {\left\langle {P_{AA}} \right\rangle_0} }\right),\label{6.6c}
\\
\begin{split}
\frac{{d{{\left\langle {{P_{GG}}} \right\rangle }_0}}}{{dt}} &= \Gamma \left( {1 + \cos {k_0}d} \right){\left\langle {{P_{SS}}} \right\rangle _0} \\
& \qquad\quad + \Gamma \left( {1 - \cos {k_0}d} \right){\left\langle {{P_{AA}}} \right\rangle _0}.\end{split}\label{6.6d}
\end{align}
\end{subequations}
For the off-diagonal matrix elements of the transition operator (which we will refer to as
coherences) we obtain:
\begin{subequations}\label{6.7}
\begin{align}
& \frac{{d{{\left\langle {{P_{GE}}} \right\rangle }_0}}}{{dt}} =  - \left( {2i\Omega  + \Gamma } \right){\left\langle {{P_{GE}}} \right\rangle _0},\label{6.7a}
\\
& \frac{{d{{\left\langle {{P_{AS}}} \right\rangle }_0}}}{{dt}} =  - \Gamma \left( {1 + i\sin {k_0}d} \right)\,{\left\langle {{P_{AS}}} \right\rangle _0},\label{6.7b}
\\
\begin{split}
& \frac{{d{{\left\langle {{P_{AE}}} \right\rangle }_0}}}{{dt}} =  - i\left( {\Omega  + \frac{\Gamma }{2}\sin {k_0}d} \right){\left\langle {{P_{AE}}} \right\rangle _0} \\
& \qquad \qquad \qquad - \frac{\Gamma }{2}\left( {3 - \cos {k_0}d} \right){\left\langle {{P_{AE}}} \right\rangle _0},\end{split}\label{6.7c}
\\
\begin{split}
& \frac{{d{{\left\langle {{P_{SE}}} \right\rangle }_0}}}{{dt}} =  - i\left( {\Omega  - \frac{\Gamma }{2}\sin {k_0}d} \right){\left\langle {{P_{SE}}} \right\rangle _0} \\
& \qquad \qquad \qquad - \frac{\Gamma }{2}\left( {3 + \cos {k_0}d} \right){\left\langle {{P_{SE}}} \right\rangle _0},\end{split}\label{6.7d}
\\
\begin{split}
& \frac{{d{{\left\langle {{P_{GA}}} \right\rangle }_0}}}{{dt}} =  - i\left( {\Omega  - \frac{\Gamma }{2}\sin {k_0}d} \right){\left\langle {{P_{GA}}} \right\rangle _0} \\
& - \Gamma \left( {1 - \cos {k_0}d} \right){\left\langle {{P_{AE}}} \right\rangle _0} - \frac{\Gamma }{2}\left( {1 - \cos {k_0}d} \right){\left\langle {{P_{GA}}} \right\rangle _0},\end{split}\label{6.7e}
\\
\begin{split}
& \frac{{d{{\left\langle {{P_{GS}}} \right\rangle }_0}}}{{dt}} =  - i\left( {\Omega  + \frac{\Gamma }{2}\sin {k_0}d} \right){\left\langle {{P_{GS}}} \right\rangle _0} \\
& + \Gamma \left( {1 + \cos {k_0}d} \right){\left\langle {{P_{SE}}} \right\rangle _0} - \frac{\Gamma }{2}\left( {1 + \cos {k_0}d} \right){\left\langle {{P_{GS}}} \right\rangle _0}.\end{split}\label{6.7f}
\end{align}
\end{subequations}
Thus in the basis (\ref{6.2}), the equations for populations are decoupled from those for
the coherences. Moreover, first four equations for the coherences are fully independent
and related only to their corresponding matrix elements. These equations can be solved
without any problems since the initial conditions, which are based on the definition of
transition operator (\ref{2.5}), are always unique: ${P_{ij}}(0) = \left|i\right\rangle
\left\langle j\right|$.

By solving two groups of equations we find all matrix elements for the transition operator
for a two-qubit system in an open waveguide. For the populations we obtain the following
solutions:
\begin{subequations}\label{6.8}
\begin{align}
&{\left\langle {{P_{EE}}(t)} \right\rangle _0} = {e^{ - 2\Gamma t}}\left| E \right\rangle \left\langle E \right|,\label{6.8a}
\\
\begin{split}
&{\left\langle {{P_{SS}}(t)} \right\rangle _0} = \left| S \right\rangle \left\langle S \right|{e^{ - {\Gamma _ + }t}}\\
& \qquad \qquad - \frac{{1 + \cos {k_0}d}}{{1 - \cos {k_0}d}}\left( {{e^{ - 2\Gamma t}} - {e^{ - {\Gamma _ + }t}}} \right)\left| E \right\rangle \left\langle E \right|,\end{split}\label{6.8b}
\\
\begin{split}
&{\left\langle {{P_{AA}}(t)} \right\rangle _0} = \left| A \right\rangle \left\langle A \right|{e^{ - {\Gamma _ - }t}}\\
& \qquad \qquad - \frac{{1 - \cos {k_0}d}}{{1 + \cos {k_0}d}}\left( {{e^{ - 2\Gamma t}} - {e^{ - {\Gamma _ - }t}}} \right)\left| E \right\rangle \left\langle E \right|,\end{split}\label{6.8c}
\\
\begin{split}
&{\left\langle {{P_{GG}}(t)} \right\rangle _0} = \left| G \right\rangle \left\langle G \right| \\
& - \left( {{e^{ - {\Gamma _ + }t}} - 1} \right)\left| S \right\rangle \left\langle S \right| - \left( {{e^{ - {\Gamma _ - }t}} - 1} \right)\left| A \right\rangle \left\langle A \right|\\
& + \frac{{{{\left( {1 + \cos {k_0}d} \right)}^2}}}{{1 - \cos {k_0}d}}\left[ {\frac{{\left( {{e^{ - 2\Gamma t}} - 1} \right)}}{2} - \frac{{\left( {{e^{ - {\Gamma _ + }t}} - 1} \right)}}{{1 + \cos {k_0}d}}} \right]\left| E \right\rangle \left\langle E \right|\\
& + \frac{{{{\left( {1 - \cos {k_0}d} \right)}^2}}}{{1 + \cos {k_0}d}}\left[ {\frac{{\left( {{e^{ - 2\Gamma t}} - 1} \right)}}{2} - \frac{{\left( {{e^{ - {\Gamma _ - }t}} - 1} \right)}}{{1 - \cos {k_0}d}}} \right]\left| E \right\rangle \left\langle E \right|.
\end{split}\label{6.8d}
\end{align}
\end{subequations}
For coherences we obtain:
\begin{subequations}\label{6.9}
\begin{align}
&{\left\langle {{P_{GE}}(t)} \right\rangle _0} = {e^{ - \left( {2i\Omega  + \Gamma } \right)t}}\left| G \right\rangle \left\langle E \right|, \label{6.9a}
\\
&{\left\langle {{P_{AS}}(t)} \right\rangle _0} = {e^{ - \Gamma \left( {1 + i\sin ({k_0}d)} \right)t}}\left| A \right\rangle \left\langle S \right|, \label{6.9b}
\\
&{\left\langle {{P_{AE}}(t)} \right\rangle _0} = {e^{ - \left( {i{\Omega _ + } + \frac{{{\Gamma _ - }}}{2} + \Gamma } \right)t}}\left| A \right\rangle \left\langle E \right|, \label{6.9c}
\\
&{\left\langle {{P_{SE}}(t)} \right\rangle _0} = {e^{ - \left( {i{\Omega _ - } + \frac{{{\Gamma _ + }}}{2} + \Gamma } \right)t}}\left| S \right\rangle \left\langle E \right|, \label{6.9d}
\\
\begin{split}
&{\left\langle {{P_{GA}}(t)} \right\rangle _0} = {e^{ - \left( {i\,{\Omega _ - } + \frac{{{\Gamma _ - }}}{2}} \right)t}}\left| G \right\rangle \left\langle A \right| + \\
& \frac{{1 - \cos {k_0}d}}{{1 + i\sin {k_0}d}}\left( {{e^{ - \left( {i{\Omega _ + } + \frac{{{\Gamma _ - }}}{2} + \Gamma } \right)t}} - {e^{ - \left( {i\,{\Omega _ - } + \frac{{{\Gamma _ - }}}{2}} \right)t}}} \right)\left| A \right\rangle \left\langle E \right|,\label{6.9e}
\end{split}
\\
\begin{split}
&{\left\langle {{P_{GS}}(t)} \right\rangle _0} = {e^{ - \left( {i\,{\Omega _ + } + \frac{{{\Gamma _ + }}}{2}} \right)t}}\left| G \right\rangle \left\langle S \right| - \\
& \frac{{1 + \cos {k_0}d}}{{1 - i\sin {k_0}d}}\left( {{e^{ - \left( {i{\Omega _ - } + \frac{{{\Gamma _ + }}}{2} + \Gamma } \right)t}} - {e^{ - \left( {i\,{\Omega _ + } + \frac{{{\Gamma _ + }}}{2}} \right)t}}} \right)\left| S \right\rangle \left\langle E \right|.\label{6.9f}
\end{split}
\end{align}
\end{subequations}
Here for simplification, we introduce the shifted resonant frequencies and modified
decay rates:
\begin{subequations}\label{6.10}
\begin{align}
{\Omega _ + } &= \Omega  + \frac{\Gamma }{2}\sin {k_0}d;\quad \quad {\Omega _ - } = \Omega  - \frac{\Gamma }{2}\sin {k_0}d;\label{6.10a}
\\
{\Gamma _ + } &= \Gamma \left( {1 + \cos {k_0}d} \right);\quad \quad {\Gamma _ - } = \Gamma \left( {1 - \cos {k_0}d} \right);\label{6.10b}
\end{align}
\end{subequations}
which depend on the effective distance between the qubits.

Unlike the usual solution for the density matrix, expressions (\ref{6.8}) and (\ref{6.9})
are the operator functions. Nevertheless, knowing the expressions for the matrix elements
of the transition operator, we can easily find the density matrix using relations
(\ref{2.12}) or (\ref{2.13}).

\section{Transition probabilities for two qubits}

As  was noted in Sec.2, the probability of a system to transit
from one state to another can be found with the aid of  transition
operators (see Eq. \ref{2.7}). Here we calculate the probabilities
which contribute to the total rate of superradiant emission which
will be given below in Sec. 7.

For both qubits initially in an excited state $\left|{{\Psi
_0}}\right\rangle=\left|{ee} \right\rangle=\left|E\right\rangle$
we can find the probability that at time $t$ the system remains in
the initial state:
\begin{equation}\label{6.11a}
{W_{E \to E}} = \left\langle E \right|{P_{EE}}\left| E \right\rangle  = {e^{ - 2\Gamma t}}.
\end{equation}


The probabilities for both qubits to decay to symmetric and
asymmetric state are as follows:

\begin{multline}\label{6.11c}
{W_{E \to S}} = \left\langle E \right|{P_{SS}}\left| E \right\rangle  = \\
 - \frac{{1 + \cos {k_0}d}}{{1 - \cos {k_0}d}}\left( {{e^{ - 2\Gamma t}} - {e^{ - \Gamma \left( {1 + \cos {k_0}d} \right)t}}} \right).
\end{multline}

\begin{multline}\label{6.11d}
{W_{E \to A}} = \left\langle E \right|{P_{AA}}\left| E \right\rangle  =  \\
- \frac{{1 - \cos {k_0}d}}{{1 + \cos {k_0}d}}\left( {{e^{ - 2\Gamma t}} - {e^{ - \Gamma \left( {1 - \cos {k_0}d} \right)t}}} \right).
\end{multline}

When qubits are initially in a symmetric, $ \left| {\Psi _0 }
\right\rangle  = {{\left( {\left| {ge} \right\rangle  + \left|
{eg} \right\rangle } \right)} \mathord{\left/
 {\vphantom {{\left( {\left| {ge} \right\rangle  - \left| {eg} \right\rangle } \right)} {\sqrt 2 }}} \right.
 \kern-\nulldelimiterspace} {\sqrt 2 }} = \left| S \right\rangle$,
 or asymmetric, $\left| {\Psi _0 } \right\rangle  =
{{\left( {\left| {ge} \right\rangle  - \left| {eg} \right\rangle }
\right)} \mathord{\left/
 {\vphantom {{\left( {\left| {ge} \right\rangle  - \left| {eg} \right\rangle } \right)} {\sqrt 2 }}} \right.
 \kern-\nulldelimiterspace} {\sqrt 2 }} = \left| A \right\rangle $
 states, the probabilities of the system to remain in the initial states are given by:

\begin{equation}\label{6.14a}
{W_{S \to S}} = \left\langle S \right|{P_{SS}}\left| S
\right\rangle  = {e^{ - \Gamma \left( {1 + \cos {k_0}d}
\right)t}}.
\end{equation}

\begin{equation}\label{6.15a}
{W_{A \to A}} = \left\langle A \right|{P_{AA}}\left| A
\right\rangle  = {e^{ - \Gamma \left( {1 - \cos {k_0}d} \right)t}}
\end{equation}

As is clear from (\ref{6.8b}) and (\ref{6.8c}) the transitions
between symmetric and asymmetric  states are forbidden:
\begin{equation}\label{6.15c}
W_{A \to S}  = \left\langle A \right|P_{SS} \left| A \right\rangle
= 0;\quad W_{S \to A}  = \left\langle S \right|P_{AA} \left| S
\right\rangle  = 0
\end{equation}
Therefore, the $|A\rangle $ and $|S\rangle $  states are
completely decoupled from each other no matter what is the value
of $k_0d$.

The symmetric and asymmetric states have different decay rates
which depend on the value of $k_0d$. As is seen from
(\ref{6.14a}), (\ref{6.15a}) for a given value of $k_0d$ the decay
rate for the state $|S\rangle $ is always greater than that for
the state $|A\rangle $. In addition, for $k_0d=2n\pi$, where $n$
is a positive integer or $0$, the population of the $|A\rangle $
remains constant ($W_{A\rightarrow A}=1$). In this case, the state
$|A\rangle $ is called the dark state since it does not interact
with the electromagnetic field, while the state $|S\rangle $ is
called a bright state. If $k_0d=(2n+1)\pi$ the situation is
reversed: the state $|S\rangle $ becomes a dark state, while the
state $|A\rangle $ becomes a bright state.

 Finally, for the calculation of
$W(t)$ we will need the off-diagonal matrix elements:
\begin{equation}\label{6.15d}
\left\langle A \right|\left\langle {P_{AS} (t)} \right\rangle _0
\left| S \right\rangle  = e^{ - \Gamma \left( {1 + i\sin (k_0 d)}
\right)t}
\end{equation}

The transitions (\ref{6.11c}, \ref{6.11d}) depend on the effective
distance between the qubits $k_0d$. For example, for $k_0d=\pi/2$
there are equal probabilities of transitions to symmetric and
asymmetric  states, $ W_{E \to S}  = W_{E \to A}  = e^{ - 2\Gamma
t}  - e^{ - \Gamma t}$.

We should separately consider the case when $k_0d=n\pi$. For this
case, both the numerator and denominator  in (\ref{6.11c}) and
(\ref{6.11d}) tend to zero. A correct solution can be obtained if
we put $k_0d = n\pi$ directly in the equations (\ref{6.6b}) and
(\ref{6.6c}), or by expanding $cos(k_0d)$ near $k_0d\approx
n\pi+\epsilon$ where $\epsilon$ is a small value. Both approaches
give the same result. For $k_0d = 2n\pi$ we obtain the following
transition probabilities:
\begin{equation}\label{6.12}
W_{E \to S}  = 2\Gamma te^{ - 2\Gamma t} ,\quad \quad W_{E \to A}
= 0
\end{equation}

As it is clearly seen from (\ref{6.12}), for an even number of $n
= 0, 2, 4\ldots$, the transition from state $\left| E
\right\rangle$ to asymmetric entangled state $\left| A
\right\rangle$ is forbidden. For an odd number of $n = 1, 3\ldots$
the situation is reversed: the transition to symmetric state is
now forbidden, and for transition to asymmetric state we get the
relation ${W_{E\to A}}=2\Gamma t{e^{- 2\Gamma t}}$.

\section{Superradiant spectra of two qubits in a waveguide}

Now we switch to the calculation of radiation spectrum for a two-qubit system. As was shown
in (\ref{4.2}), the spectrum can be found using a set of atomic correlation functions. For
$N = 2$ we obtain:
\begin{subequations}\label{6.18}
\begin{align}
&\left\langle {a_k^\dag (t){a_k}(t)} \right\rangle  = {\left\langle {a_k^\dag (t){a_k}(t)} \right\rangle _ + } + {\left\langle {a_k^\dag (t){a_k}(t)} \right\rangle _ - },\label{6.18a} \\
&\begin{gathered} {\left\langle {a_k^\dag (t){a_k}(t)}
\right\rangle _ + } = {\left| {{g_k}} \right|^2}\int\limits_0^t
{d\tau } \int\limits_0^\tau  {d\tau '} {e^{ - i\omega (\tau  -
\tau ')}}\Theta_k (\tau ,\tau '),\label{6.18b}
\end{gathered}
\\
&\begin{gathered} {\left\langle {a_k^\dag (t){a_k}(t)}
\right\rangle _ - } = {\left| {{g_k}} \right|^2}\int\limits_0^t
{d\tau } \int\limits_\tau ^t {d\tau '} {e^{ - i\omega (\tau  -
\tau ')}}\Theta_k (\tau ,\tau ').\label{6.18c}
\end{gathered}
\end{align}
\end{subequations}
where:
\begin{multline}
\Theta_k (\tau ,\tau ') = \left\langle {{\sigma _ + }^{(1)}(\tau
){\sigma _ - }^{(1)}(\tau ')} \right\rangle  + \left\langle
{{\sigma _ + }^{(2)}(\tau ){\sigma _ - }^{(2)}(\tau ')}
\right\rangle  \\+ {e^{ - ikd}}\left\langle {{\sigma _ +
}^{(1)}(\tau ){\sigma _ - }^{(2)}(\tau ')} \right\rangle  +
{e^{ikd}}\left\langle {{\sigma _ + }^{(2)}(\tau ){\sigma _ -
}^{(1)}(\tau ')} \right\rangle.
\end{multline}

In order to correctly calculate the two-time spin correlation
functions we subdivide the whole spectrum (\ref{6.18a}) into two
parts, for positive time difference (\ref{6.18b}) (when
$\tau>\tau^\prime$) and negative time difference (\ref{6.18c})
(when $\tau<\tau^\prime$).

We want to remind again that in (\ref{6.18}) the positive
$k=+\omega/v_g$ corresponds to a forward, right moving wave, while
the negative $k=-\omega/v_g$ corresponds to a backward, left
moving wave,

To find complete spectra (\ref{6.18a}) one should calculate the
two-time correlation functions using (\ref{4.13}). From
(\ref{6.3}) we can express the lowering and raising spin operators
in terms of basis set (\ref{6.2}):
\begin{subequations}\label{6.19}
\begin{align}
\sigma _ + ^{(1,2)} =  & \frac{1}{{\sqrt 2 }}\left( {\left| S \right\rangle \left\langle G \right| \mp \left| A \right\rangle \left\langle G \right| + \left| E \right\rangle \left\langle S \right| \pm \left| E \right\rangle \left\langle A \right|} \right),\label{6.19a} \\
\sigma _ - ^{(1,2)} =  & \frac{1}{{\sqrt 2 }}\left( {\left| G \right\rangle \left\langle S \right| \mp \left| G \right\rangle \left\langle A \right| + \left| S \right\rangle \left\langle E \right| \pm \left| A \right\rangle \left\langle E \right|} \right),\label{6.19b}
\end{align}
\end{subequations}
and by switching to a Heisenberg picture, we find:
\begin{subequations}\label{6.20}
\begin{align}
\sigma _ + ^{(1,2)}(t) =  & \frac{1}{{\sqrt 2 }}\left( {{P_{SG}}(t) \mp {P_{AG}}(t) + {P_{ES}}(t) \pm {P_{EA}}(t)} \right),\label{6.20a} \\
\sigma _ - ^{(1,2)}(t) =  & \frac{1}{{\sqrt 2 }}\left( {{P_{GS}}(t) \mp {P_{GA}}(t) + {P_{SE}}(t) \pm {P_{AE}}(t)} \right),\label{6.20b}
\end{align}
\end{subequations}
Thus, one can find the complete spectra (\ref{6.18a}) by calculating four two-time
correlation functions with the already obtained transition operators (\ref{6.8}) and (\ref{6.9}).

\begin{widetext}
For the positive time difference $\tau>\tau^\prime$ we obtain the
following general expression:
\begin{equation}\label{6.21}
\begin{gathered}
  \left\langle {a_{k}^\dag  (t)a_{k} (t)} \right\rangle _ +   = \left| {g_{k} } \right|^2 \int\limits_0^t {d\tau } \int\limits_0^\tau {d\tau '} e^{ - i\omega (\tau  - \tau ')}  \\
 \times \left[ {} \right.\left( {1 + \cos (kd)} \right)\left( {\left\langle E \right|\left\langle {P_{SG} (\tau  - \tau ')} \right\rangle _0 \left| S \right\rangle } \right.\left\langle E \right|\left\langle {P_{EE} (\tau ')} \right\rangle _0 \left| E \right\rangle  + \left\langle E \right|\left\langle {P_{ES} (\tau  - \tau ')} \right\rangle _0 \left| S \right\rangle \left\langle E \right|\left\langle {P_{EE} (\tau ')} \right\rangle _0 \left| E \right\rangle  \\
   + \left. {\left\langle S \right|\left\langle {P_{SG} (\tau  - \tau ')} \right\rangle _0 \left| G \right\rangle \left\langle E \right|\left\langle {P_{SS} (\tau ')} \right\rangle _0 \left| E \right\rangle } \right)\left\langle E \right|\rho _S (0)\left| E \right\rangle  \\
  +\left( {1 - \cos (kd)} \right)\left( {\left\langle E \right|\left\langle {P_{EA} (\tau  - \tau ')} \right\rangle _0 \left| A \right\rangle } \right.\left\langle E \right|\left\langle {P_{EE} (\tau ')} \right\rangle _0 \left| E \right\rangle  - \left\langle E \right|\left\langle {P_{AG} (\tau  - \tau ')} \right\rangle _0 \left| A \right\rangle \left\langle E \right|\left\langle {P_{EE} (\tau ')} \right\rangle _0 \left| E \right\rangle  \\
   + \left. {\left\langle A \right|\left\langle {P_{AG} (\tau  - \tau ')} \right\rangle _0 \left| G \right\rangle \left\langle E \right|\left\langle {P_{AA} (\tau ')} \right\rangle _0 \left| E \right\rangle } \right)\left\langle E \right|\rho _S (0)\left| E \right\rangle  \\
   + \left( {1 + \cos (kd)} \right)\left\langle S \right|\left\langle {P_{SG} (\tau  - \tau ')} \right\rangle _0 \left| G \right\rangle \left\langle S \right|\left\langle {P_{SS} (\tau ')} \right\rangle _0 \left| S \right\rangle \left\langle S \right|\rho _S (0)\left| S \right\rangle  \\
   + \left( {1 - \cos (kd)} \right)\left\langle A \right|\left\langle {P_{AG} (\tau  - \tau ')} \right\rangle _0 \left| G \right\rangle \left\langle A \right|\left\langle {P_{AA} (\tau ')} \right\rangle _0 \left| A \right\rangle \left\langle A \right|\rho _S (0)\left| A \right\rangle  \\
   + i\sin (kd)\left\langle A \right|\left\langle {P_{AG} (\tau  - \tau ')} \right\rangle _0 \left| G \right\rangle \left\langle A \right|\left\langle {P_{AS} (\tau ')} \right\rangle _0 \left| S \right\rangle \left\langle S \right|\rho _S (0)\left| A \right\rangle  \\
   - i\sin (kd)\left\langle S \right|\left\langle {P_{SG} (\tau  - \tau ')} \right\rangle _0 \left| G \right\rangle \left\langle S \right|\left\langle {P_{SA} (\tau ')} \right\rangle _0 \left| A \right\rangle \left. {\left\langle A \right|\rho _S (0)\left| S \right\rangle } \right]\left. {} \right]; \\
\end{gathered}
\end{equation}
and for the negative time difference $\tau<\tau^\prime$ we obtain:
\begin{equation}\label{6.21b}
\begin{gathered}
  \left\langle {a_{k}^\dag  (t)a_{k} (t)} \right\rangle _ -   = \left| {g_{k} } \right|^2 \int\limits_0^t {d\tau } \int\limits_\tau ^t {d\tau '} e^{ - i\omega (\tau  - \tau ')}  \\
   \times \left[ {} \right.\left( {1 + \cos (k d)} \right)\left( {\left\langle S \right|\left\langle {P_{GS} (\tau ' - \tau )} \right\rangle _0 \left| E \right\rangle \left\langle E \right|\left\langle {P_{EE} (\tau )} \right\rangle _0 \left| E \right\rangle  + \left\langle S \right|\left\langle {P_{SE} (\tau ' - \tau )} \right\rangle _0 \left| E \right\rangle \left\langle E \right|\left\langle {P_{EE} (\tau )} \right\rangle _0 \left| E \right\rangle } \right.   \\
  \left. { + \left\langle G \right|\left\langle {P_{GS} (\tau ' - \tau )} \right\rangle _0 \left| S \right\rangle \left\langle E \right|\left\langle {P_{SS} (\tau )} \right\rangle _0 \left| E \right\rangle } \right)\left\langle E \right|\rho _S (0)\left| E \right\rangle  \\
  +\left( {1 - \cos (k d)} \right)\left( {\left\langle A \right|\left\langle {P_{AE} (\tau ' - \tau )} \right\rangle _0 \left| E \right\rangle \left\langle E \right|\left\langle {P_{EE} (\tau )} \right\rangle _0 \left| E \right\rangle  - \left\langle A \right|\left\langle {P_{GA} (\tau ' - \tau )} \right\rangle _0 \left| E \right\rangle \left\langle E \right|\left\langle {P_{EE} (\tau )} \right\rangle _0 \left| E \right\rangle } \right. \\
  \left. { + \left\langle G \right|\left\langle {P_{GA} (\tau ' - \tau )} \right\rangle _0 \left| A \right\rangle \left\langle E \right|\left\langle {P_{AA} (\tau )} \right\rangle _0 \left| E \right\rangle } \right)\left\langle E \right|\rho _S (0)\left| E \right\rangle  \\
   + \left( {1 + \cos (k d)} \right)\left\langle G \right|\left\langle {P_{GS} (\tau ' - \tau )} \right\rangle _0 \left| S \right\rangle \left\langle S \right|\left\langle {P_{SS} (\tau )} \right\rangle _0 \left| S \right\rangle \left\langle S \right|\rho _S (0)\left| S \right\rangle  \\
   + \left( {1 - \cos (k d)} \right)\left\langle G \right|\left\langle {P_{GA} (\tau ' - \tau )} \right\rangle _0 \left| A \right\rangle \left\langle A \right|\left\langle {P_{AA} (\tau )} \right\rangle _0 \left| A \right\rangle \left\langle A \right|\rho _S (0)\left| A \right\rangle  \\
   - i\sin (k d)\left\langle G \right|\left\langle {P_{GA} (\tau ' - \tau )} \right\rangle _0 \left| A \right\rangle \left\langle S \right|\left\langle {P_{SA} (\tau )} \right\rangle _0 \left| A \right\rangle \left\langle A \right|\rho _S (0)\left| S \right\rangle  \\
   + i\sin (k d)\left\langle G \right|\left\langle {P_{GS} (\tau ' - \tau )} \right\rangle _0 \left| S \right\rangle \left\langle A \right|\left\langle {P_{AS} (\tau )} \right\rangle _0 \left| S \right\rangle \left\langle S \right|\rho _S (0)\left| A \right\rangle \left. {} \right] \\
\end{gathered}
\end{equation}
\end{widetext}
As is seen from (\ref{6.21}), (\ref{6.21b}) only the initial
density matrix of the form  $\rho_S(0)=a|E \rangle \langle E|+b|A
\rangle \langle A|+c|S \rangle \langle S|+d|A \rangle \langle
S|+f|S \rangle \langle A|$, where $a,b,c,d,f$ are arbitrary
complex values, contributes to the radiation spectrum.

We can also calculate  a total emission rate (\ref{4.2a}) for
$N=2$:


\begin{multline}\label{6.22a}
W(t) = \frac{\Gamma }{2}\left( {\left\langle {{\sigma _ + }^{(1)}(t){\sigma _ - }^{(1)}(t)} \right\rangle  + \left\langle {{\sigma _ + }^{(2)}(t){\sigma _ - }^{(2)}(t)} \right\rangle } \right. \\
+\left. {{e^{ - i{k}d}}\left\langle {{\sigma _ + }^{(1)}(t){\sigma
_ - }^{(2)}(t)} \right\rangle  + {e^{i{k}d}}\left\langle {{\sigma
_ + }^{(2)}(t){\sigma _ - }^{(1)}(t)} \right\rangle } \right)
\end{multline}
where $\left\langle {{\sigma _ + }^{(i)}(t){\sigma _ - }^{(j)}(t)}
\right\rangle $ can be found  with $\tau =\tau'=t$ in either of
equations (\ref{4.13}). Thus, the emission rate can easily be
calculated since it is proportional only to single-time
correlation functions. With the help of expressions for spin
operators (\ref{6.20}) we can express (\ref{6.22a}) in terms of
transition operators:

\begin{multline}\label{6.22b}
W(t) = \left( {\Gamma \left\langle E \right|{{\left\langle {{P_{EE}}(t)} \right\rangle }_0}\left| E \right\rangle  + \frac{{{\Gamma _ + }}}{2}\left\langle E \right|{{\left\langle {{P_{SS}}(t)} \right\rangle }_0}\left| E \right\rangle } \right.\\
\left. { + \frac{{{\Gamma _ - }}}{2}\left\langle E \right|{{\left\langle {{P_{AA}}(t)} \right\rangle }_0}\left| E \right\rangle } \right)\left\langle E \right|{\rho _S}(0)\left| E \right\rangle \\
 + \frac{{{\Gamma _ + }}}{2}\left\langle S \right|{\left\langle {{P_{SS}}(t)} \right\rangle _0}\left| S \right\rangle \left\langle S \right|{\rho _S}(0)\left| S \right\rangle \\
 + \frac{{{\Gamma _ - }}}{2}\left\langle A \right|{\left\langle {{P_{AA}}(t)} \right\rangle _0}\left| A \right\rangle \left\langle A \right|{\rho _S}(0)\left| A \right\rangle \\
 + i\frac{\Gamma }{2}\sin ({k}d)\left\langle A \right|{\left\langle {{P_{AS}}(t)} \right\rangle _0}\left| S \right\rangle \left\langle S \right|{\rho _S}(0)\left| A \right\rangle \\
 - i\frac{\Gamma }{2}\sin ({k}d)\left\langle S \right|{\left\langle {{P_{SA}}(t)} \right\rangle _0}\left| A \right\rangle \left\langle A \right|{\rho _S}(0)\left| S \right\rangle
\end{multline}

The expressions (\ref{6.21}), (\ref{6.21b}), and (\ref{6.22b}) are
the central result that we use in the following to calculate  the
super- and subradiant spectra and emission rates in two-qubit
system for various initial configuration. They can be applied to
any initial density matrix $\rho_S(0)$.

Below we consider several excited configurations of the two-qubit
system. For every configuration we calculate the emission photon
spectrum and the total rate of photon emission for different
values of $k_0d$. In order to make evident the influence of the
second qubit on the radiation spectrum, we compare these
quantities with those for a single qubit in the system.

In all figures to this section the radiation spectral densities
and the emission rates are given in dimensionless units
${S}(\omega)2L\Omega/v_g$ and $W(t)/\Gamma$, respectively. All
calculations are made for $\Gamma/\Omega=0.05$.

\subsection{Initial symmetric and asymmetric states}

We start with initially prepared entangled states in the form of a
symmetrical state
$\left|{\Psi(0)}\right\rangle{=\left({\left|{{e_1}{g_2}}\right\rangle
+\left|{{g_1}{e_2}} \right\rangle}\right)}/{\sqrt
2}=\left|S\right\rangle$ and an asymmetrical state
$\left|{\Psi(0)}\right\rangle{=\left({\left|{{g_1}{e_2}}\right\rangle
-\left|{{e_1}{g_2}} \right\rangle}\right)}/{\sqrt 2}= \left| A
\right\rangle$. The experimental technique for the preparation of
these entangled states is widely known in the circuit QED field
and can be implemented by the sequence of Hadamard and CNOT gates
\cite{Wendin17}.

As the qubit-photon coupling is efficient at
$\omega\approx\Omega$, we perform subsequent calculations for
$k\approx\pm k_0$ where $k_0=\Omega/v_g$.

As is seen from (\ref{6.21}), (\ref{6.21b}) the contribution of
the symmetric  and asymmetric initial states,
${\rho_S}(0)=\left|S\right\rangle\left\langle S \right|$,
${\rho_S}(0)=\left|A\right\rangle\left\langle A \right|$ are the
even function of $k$. Therefore, the corresponding spectra are the
same in both directions. From (\ref{6.21}), (\ref{6.21b}), and
(\ref{6.8b}, \ref{6.9f}) we obtain:

\begin{multline}\label{6.27a}
{\left\langle {a_k^\dag (t){a_k}(t)} \right\rangle _S} = \\
\frac{{{\upsilon _g}{\Gamma _ + }}}{{2L}}\frac{{\left( {{e^{\left( {i{\delta _ + } - \frac{{{\Gamma _ + }}}{2}} \right)t}} - 1} \right)\left( {{e^{ - \left( {i{\delta _ + } + \frac{{{\Gamma _ + }}}{2}} \right)t}} - 1} \right)}}{{\delta _ + ^2 + \frac{{\Gamma _ + ^2}}{4}}}.
\end{multline}
where we introduce the detuning parameters:
\begin{equation}\label{6.26}
{\delta _ + } = \omega  - {\Omega _ + };\quad \quad {\delta _ - }
= \omega  - {\Omega _ - };
\end{equation}

If now we let time tend to infinity, $t\to\infty$, we get a
radiation spectrum that is dependent only on the frequency:
\begin{equation}\label{6.27b}
{S_S}(\omega) = \frac{\upsilon _g}{2L}\frac{\Gamma_+}{\left({\delta_+^2 + {\Gamma _+^2}/4} \right)}
\end{equation}
For the total emission rate we obtain from (\ref{6.22b}) :

\begin{equation}\label{6.27c}
{W_S}(t) = \frac{{{\Gamma _ + }}}{2}{e^{ - {\Gamma _ + }t}}
\end{equation}

The calculation procedure for asymmetric state
${\rho_S}(0)=\left|A\right\rangle \left\langle A \right|$ is very
similar. For this initial state, we obtain for the spectrum,
spectral density, and emission rate, respectively:

\begin{multline}\label{6.29a}
{\left\langle {a_{_k}^\dag (t){a_k}(t)} \right\rangle _A} = \\
\frac{{{\upsilon _g}{\Gamma _ - }}}{{2L}}\frac{{\left( {{e^{\left( {i{\delta _ - } - \frac{{{\Gamma _ - }}}{2}} \right)t}} - 1} \right)\left( {{e^{ - \left( {i{\delta _ - } + \frac{{{\Gamma _ - }}}{2}} \right)t}} - 1} \right)}}{{\delta _ - ^2 + \frac{{\Gamma _ - ^2}}{4}}}
\end{multline}

\begin{equation}\label{6.29b}
{S_A}(\omega) = \frac{\upsilon _g}{2L}\frac{\Gamma_-}{\left({\delta_-^2 + {\Gamma _-^2}/4} \right)}
\end{equation}

\begin{equation}\label{6.29c}
{W_A}(t) = \frac{{{\Gamma_-}}}{2}{e^{-{\Gamma_-}t}}
\end{equation}

Below we compare these quantities with those for an initially
excited single qubit with the same frequency $\Omega$ and the
decay rate $\Gamma$. For this case, the spectral density and the
total emission rate are as follows:

\begin{equation}\label{5.17}
S_1(\omega ) = \frac{{\upsilon _g }} {{2L}}\frac{\Gamma } {{\left(
{\omega  - \Omega } \right)^2  + {{\Gamma ^2 } \mathord{\left/
 {\vphantom {{\Gamma ^2 } 4}} \right.
 \kern-\nulldelimiterspace} 4}}}
\end{equation}

\begin{equation}\label{5.19}
W_1(t) = \frac{\Gamma } {2}e^{ - \Gamma t}
\end{equation}

As is seen from (\ref{6.27b}) and (\ref{6.29b}) both spectra are
Lorentzian lines whose central frequencies and the widths depend
on $k_0d$. The total emission rates (\ref{6.27c}) and
(\ref{6.29c}) also depends on $k_0d$. For Dicke case, $k_0d=0$, we
obtain $W_A=0$, $W_S=\Gamma e^{-2\Gamma t}$. The comparison of
this result with (\ref{5.19}) shows that the mere presence of a
second unexcited qubit significantly alters the photon emission
from excited qubit: its initial amplitude is twice as large as
that for a single qubit and its decay proceeds at a twofold rate.
This phenomenon is called a single photon superradiance
\cite{Dicke54, Scully2009} that can occur when a single-photon
Dicke state is formed: $N$ identical two level atoms are in a
symmetrical superposition of states with one excited atom and $N -
1$ atoms in the ground state. In this case, the decay rate of a
single photon is also equal to $N\Gamma$. The total radiated
energy must be the same for both cases: $\int_0^\infty  {W_1
(t)dt}  = \int_0^\infty  {W_S (t)dt}$.

In relation to our problem it is important to note that contrary
to free space in a one-dimensional geometry the Dicke case
$k_0d=0$ occurs also for any $k_0d=2n\pi$, where $n$ is positive
integer.

\begin{figure}
  \includegraphics[width=8 cm]{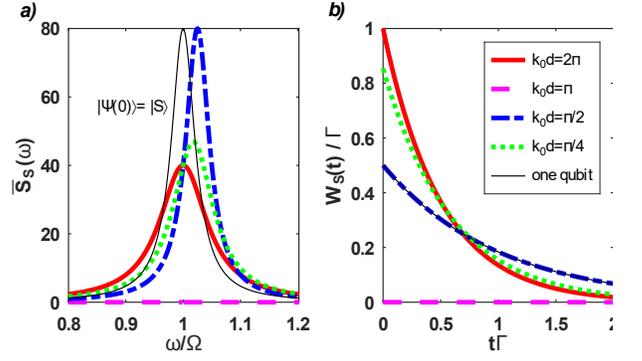}\\
  \caption {a) Radiation spectra
  $\overline{S}_{S}(\omega)={S}_{S}(\omega)2L\Omega/v_g$, expression
  (\ref{6.27b}),
  and b) Time dependence of the photon emission rate $W_S/\Gamma$,
  expression (\ref{6.27c})),
  for initial symmetric state
$|S\rangle$. For the comparison, the one-qubit case is shown by
thin solid line; $\Gamma/\Omega = 0.05$.} \label{fig1}
\end{figure}

\begin{figure}
  \includegraphics[width=8 cm]{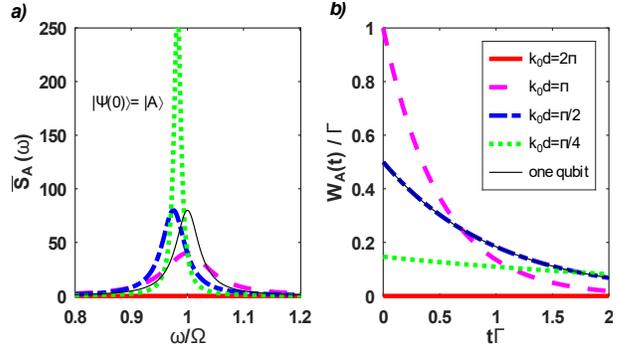}\\
  \caption {a) Radiation spectra $\overline{S}_{A}(\omega)={S}_{A}(\omega)2L\Omega/v_g$, expression
  (\ref{6.29b}), and b) Time dependence of the photon emission rate $W_A/\Gamma$, expression (\ref{6.29c})), for initial asymmetric state $|A\rangle$. For the comparison, the one-qubit case is shown by thin solid line; $\Gamma/\Omega = 0.05$.} \label{fig2}
\end{figure}

The radiation spectra for initially  states $|S\rangle$ and
$|A\rangle$, together with the corresponding values of photon
emission rates, $W_S$, $W_A$ for different values of $k_0d$ are
shown in Fig. \ref{fig1},  Fig. \ref{fig2}, respectively.

From (\ref{6.27c}) we see that for $k_0d = 2\pi$ a collective
decay rate $\Gamma_+$ becomes equal to doubled decay rate of a
single qubit: $\Gamma_+ = 2\Gamma$. This is shown by solid red
line in Fig. \ref{fig1}b. The initial intensity of the photon
emission is twice as large as that for a single qubit. The
spectral line (solid red line in Fig. \ref{fig1}a) is not shifted.
Its spectrum $S_S(\omega)$ is similar to the one for a single
qubit, but with a doubled line width and twofold decrease in the
peak value. This is, in fact, is the manifestation of the Dicke
superradiance, when the spectral line width is proportional to the
number of atoms in the system, $N = 2$ in our case. For
$k_0d=\pi/2$ the emission rates for two- and one-qubit systems are
the same; dashed blue line in (Fig.\ref{fig1}b) is superimposed on
a single qubit line. Their spectral lines are identical but are
shifted by $\Gamma/2$. If $k_0d=\pi$ the decay rate $\Gamma_+$
becomes zero, so that the symmetric state $|S\rangle$ does not
radiate at all. Therefore, we may expect that in the vicinity of
this value there exist a range of subradiant states with
$\Gamma_{sub}\ll\Gamma$.

A different picture for the decay of initially asymmetric state
$|A\rangle$ is shown in Fig.\ref{fig2}.  The superradiant emission
is seen for $k_0d=\pi$ (purple dashed line in Fig.\ref{fig2}b).
For $k_0d=\pi/2$ the decay lines  of emission rates for two- and
single-qubit systems are superimposed (Fig.\ref{fig2}b). The
spectral line of a two-qubit system is identical to the one for a
single qubit but is shifted to the left by $\Gamma/2$. A
distinctive manifestation of the subradiant decay of asymmetric
state is seen for $k_0d=\pi/4$ (green dashed line in
Fig.\ref{fig2}b). This decay is noticeably slower than the
superrradiant decay (purple dashed line in Fig.\ref{fig2}b). The
width of its spectral line is much smaller than that of a single
qubit (purple dashed line in Fig.\ref{fig2}a). If $k_0d=2\pi$ the
state $|A\rangle$ does not radiate. Here we also may expect the
range of subradiant states with $\Gamma_{sub}\ll\Gamma$ in the
vicinity of $k_0d=2\pi$.

The only common feature of the decay of the states $|S\rangle$ and
$|A\rangle$ is observed for $k_0d = (n+1/2)\pi$ which corresponds
to $\lambda_0=2d/(n+1/2)$, where $\lambda_0=2\pi\Omega/v_g$. For
this case, the radiating spectra for both symmetric and asymmetric
states have the same linewidth ${\Gamma _+} = {\Gamma_-} =\Gamma$,
but their  peaks are shifted in opposite directions because of the
frequency shift ${\delta_\pm } =\omega-(\Omega\pm\Gamma/2)$. The
evolution of their decay rates coincides with that for a single
qubit.

\begin{figure}
  \includegraphics[width=8cm]{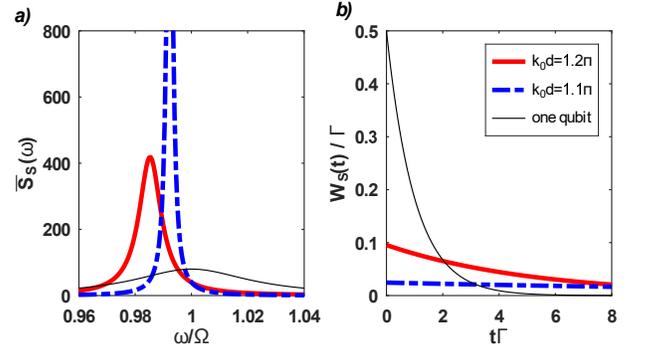}\\
  \caption{Two subradiant decays of the initial $|S\rangle$ state for $k_0d=1.2\pi$ (solid red line) and $k_0d=1.1\pi$ (dashed blue line). (a) Radiation spectra $\overline{S}_{S}(\omega)={S}_{S}(\omega)2L\Omega/v_g$; (b) Photon emission decay rate $W_S/\Gamma$. A single-qubit case is shown by thin solid line; $\Gamma/\Omega = 0.05$.}\label{fig3}
\end{figure}

We see from the Fig. \ref{fig1}, Fig. \ref{fig2} that there are
many subradiant states in the vicinity of $k_0d=\pi$ and
$k_0d=2\pi$  for initial $|S\rangle$ and $|A\rangle$ states,
respectively. As the example, two subradiant states are shown in
Fig. \ref{fig3} for initial state $|S\rangle$. The widths of the
emission spectra in Fig. \ref{fig3}a are equal to the
corresponding decay rates in Fig. \ref{fig3}b.

\subsection{Initial state with one excited qubit}

Now we consider the initial state when only the first qubit is
excited $\left| {\Psi(0)} \right\rangle  =
\left|{{e_1}{g_2}}\right\rangle$. The corresponding initial
density matrix is given by:
\begin{equation}\label{6.30}
{\rho _S}(0) = \left| {eg} \right\rangle \left\langle {eg} \right| = \frac{1}{2}\left| {S - A} \right\rangle \left\langle {S - A} \right|
\end{equation}
From (\ref{6.30}) it is seen, that two terms proportional to
$\left|S\right\rangle \left\langle S\right|$ and
$\left|A\right\rangle \left\langle A\right|$ provide the same
result we calculated in the previous section. Hence, we only need
to find the contribution of off-diagonal elements of $\rho(0)$.

Using (\ref{6.21}), (\ref{6.21b}), and explicit expressions for
the transition operators (\ref{6.8a})-(\ref{6.8d}),
(\ref{6.9a})-(\ref{6.9f}) we obtain the following expression:
\begin{multline}\label{6.32a}
{\left\langle {a_k^\dag (t){a_k}(t)} \right\rangle _{eg}} = \frac{1}{2}{\left\langle {a_k^\dag (t){a_k}(t)} \right\rangle _S} + \frac{1}{2}{\left\langle {a_k^\dag (t){a_k}(t)} \right\rangle _A}\\
 + i\frac{{{\upsilon _g}\Gamma }}{{4L}}\sin ({k}d)\left[ {\frac{{\left( {{e^{ - \left( {i{\delta _ - } + {\Gamma _ - }/2} \right)t}} - 1} \right)\left( {{e^{\left( {i{\delta _ + } - {\Gamma _ + }/2} \right)t}} - 1} \right)}}{{\left( {i{\delta _ - } + {\Gamma _ - }/2} \right)\left( {i{\delta _ + } - {\Gamma _ + }/2} \right)}}} \right.\\
\left. { - \frac{{\left( {{e^{ - \left( {i{\delta _ + } + {\Gamma _ + }/2} \right)t}} - 1} \right)\left( {{e^{\left( {i{\delta _ - } - {\Gamma _ - }/2} \right)t}} - 1} \right)}}{{\left( {i{\delta _ + } + {\Gamma _ + }/2} \right)\left( {i{\delta _ - } - {\Gamma _ - }/2} \right)}}} \right]
\end{multline}
where the first two terms are given in (\ref{6.27a}) and (\ref{6.29a}). In the limit
$t\to\infty$, we find frequency-dependent spectrum density:
\begin{multline}\label{6.32b}
{S_{eg}}(\omega ) = \frac{1}{2}{S_S}(\omega ) + \frac{1}{2}{S_A}(\omega )\\
 - \frac{{{\upsilon _g}\Gamma }}{{4L}}\frac{{\sin ({k}d)\left( {{\delta _ - }{\Gamma _ + } - {\delta _ + }{\Gamma _ - }} \right)}}{{\left( {\delta _ + ^2 + \Gamma _ + ^2/4} \right)\left( {\delta _ - ^2 + \Gamma _ - ^2/4} \right)}}
\end{multline}
where $S_S(\omega)$ and $S_A(\omega)$ are given in (\ref{6.27b}) and (\ref{6.29b}).
Finally, for the total emission rate we obtain:
\begin{multline}\label{6.32c}
{W_{eg}}(t) = \frac{{{\Gamma _ + }}}{4}{e^{ - {\Gamma _ + }t}} + \frac{{{\Gamma _ - }}}{4}{e^{ - {\Gamma _ - }t}}\\
 - \frac{{\Gamma \sin ({k}d)}}{2}{e^{ - \Gamma t}}\sin \left( {\Gamma \sin ({k_0}d)\,t} \right)
\end{multline}

In the expressions (\ref{6.32a}), (\ref{6.32b}), and (\ref{6.32c})
first two terms correspond to the contribution from the states
$|S\rangle$ and $|A\rangle$, while the second term results from
the contribution of the off-diagonal matrix elements of the
transition operator (last two lines in equations (\ref{6.21}),
(\ref{6.21b}), (\ref{6.22b}). The backward and forward radiation
corresponds to $k\approx -k_0$ and $k\approx +k_0$, respectively.

\begin{figure}
     \includegraphics[width=8 cm]{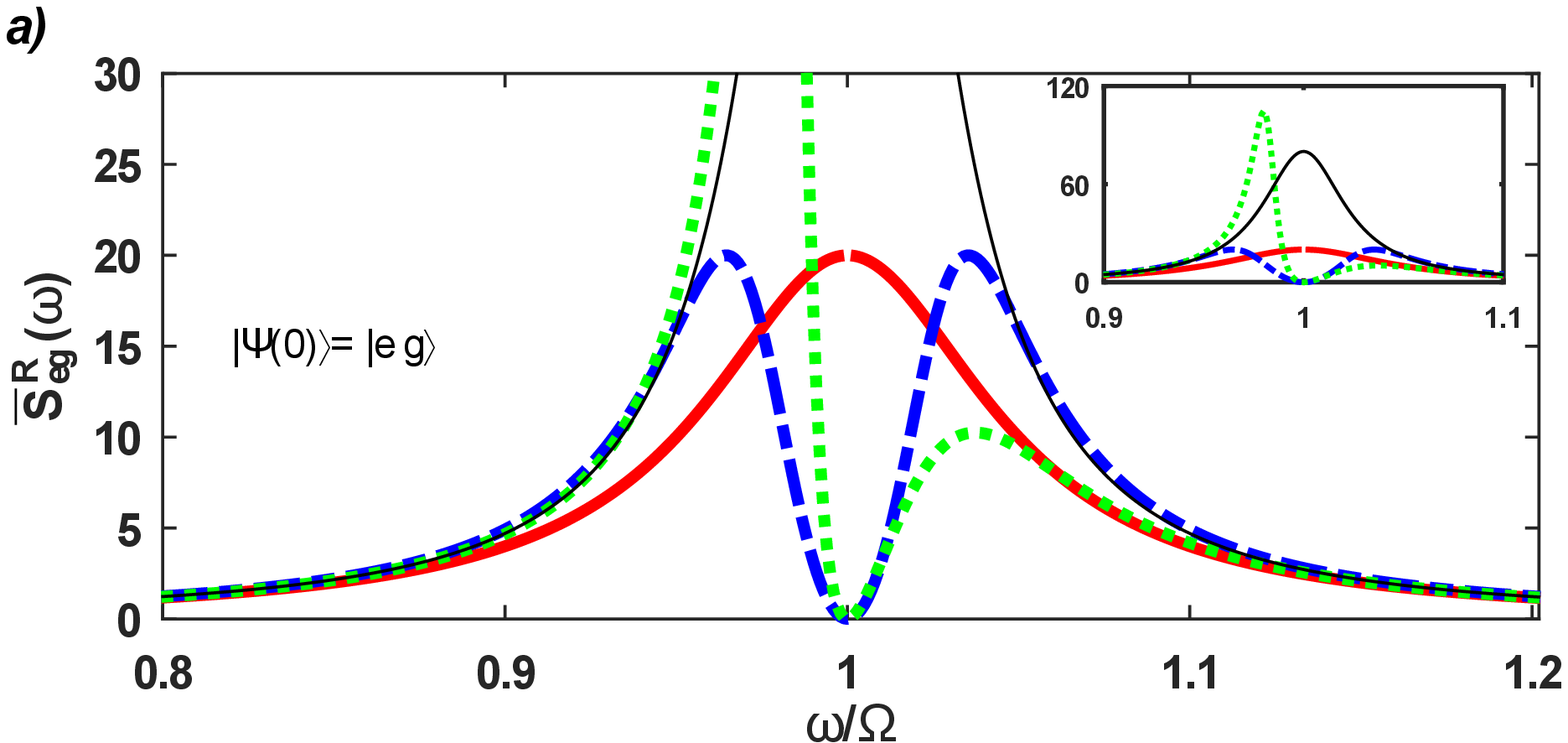}\\
     \includegraphics[width=8 cm]{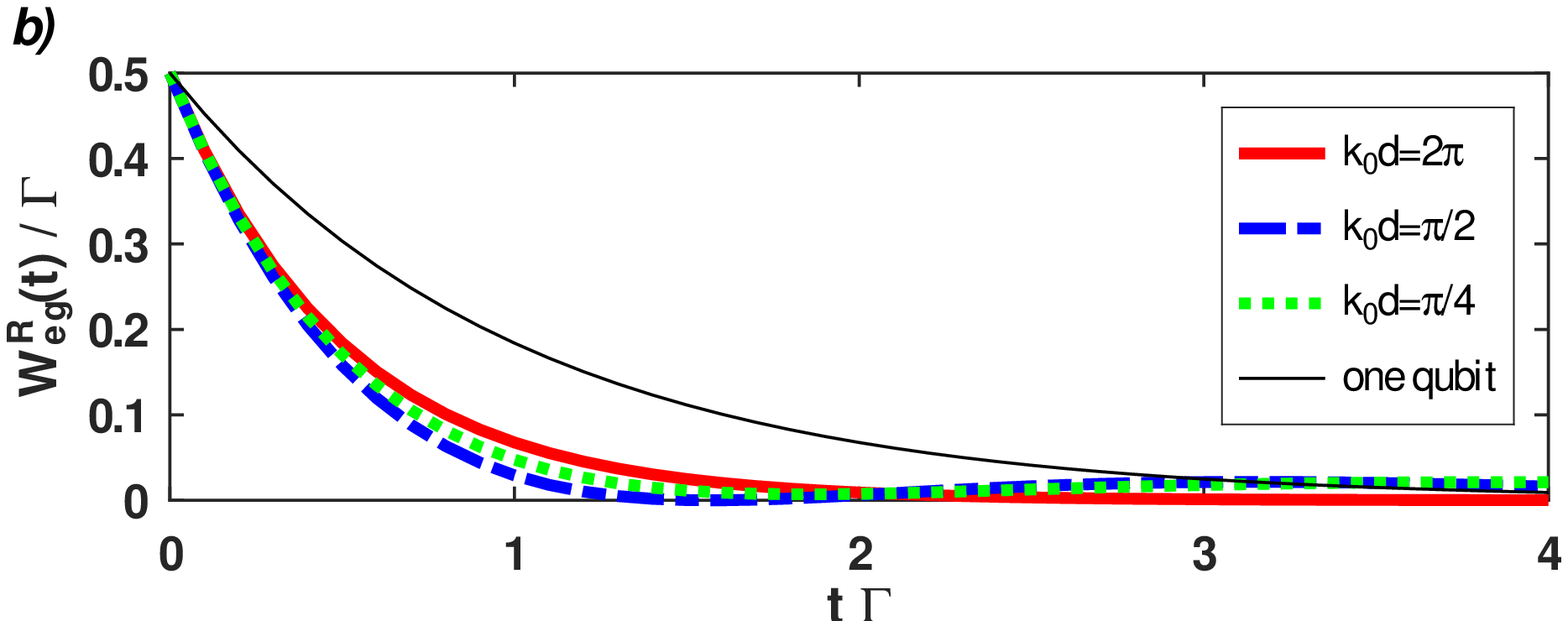}\\
 \caption{a) The forward radiation spectra $\overline{S}_{eg}^L(\omega)={S}^L_{eg}(\omega)2L\Omega/v_g$ for initially excited first qubit; b) The forward emission  decay rate, $W^L_{eg}/\Gamma$ for initially excited first qubit; For comparison a single-qubit case is shown by black thin line. $\Gamma/\Omega = 0.05$.}
  \label{fig4}
\end{figure}

The forward radiation spectra and the forward emission decay rates
are shown in Fig.\ref{fig4}a and Fig.\ref{fig4}b. Here, the
interference terms in equations (\ref{6.32b}), and (\ref{6.32c})
significantly alter the picture. If $k_0d$ is integer multiple of
$\pi$, then the interference term is zero. Therefore, for
$k_0d=2\pi$ there is a superradiant state (red, solid line in
Fig.\ref{fig4}a and b) with the decay rate being
equal to 2$\Gamma$, $W^L_{eg}(t)=0.5\Gamma e^{-2\Gamma t}$. In
this case, the spectral line is a Loretzian as shown in Fig.
\ref{fig4}a. This effect which is known  as single-atom
\cite{Dicke54} or single-photon \cite{Scully2009} superradiance
predicts the decay of excited atom at an enhanced rate in the
presence of a second atom even though that second atom is in its
ground state.

However, for other values of $k_0d$ for which the interference
term is not zero, the decay rate may also be as fast as the
superradiance decay for $k_0d=2\pi$. For example, for $k_0d=\pi/2$
we obtain from (\ref{6.32c}) $W^L_{eg}(t)=0.5\Gamma e^{-\Gamma
t}(1-\sin(\Gamma t))$ (dashed blue line in Fig. \ref{fig4}b).
Therefore, with respect to a single qubit case, all decay plots in
Fig. \ref{fig4}b within the initial time scale, $0<\Gamma t<1$,
may be considered as superradiant ones. In addition, the spectral
line for $k_0d=\pi/2$, dashed blue line in Fig. \ref{fig4}a, has a
double peak symmetrical structure. A distance between the peaks is
a measure of the coherent exchange interaction between qubits
mediated by the continuum spectra of virtual photons. For this
case, the inter-peak distance, which is determined by numerics, is
$1.45\Gamma$. By taking intermediate values of $k_0d$ we can break
this symmetry of interaction between the qubits. The plot of such
asymmetric structure, which can be a signature of Fano resonance,
is shown in Fig. \ref{fig4}a for $k_0d=\pi/4$. It is also worth
mentioning the absence of forward radiation at the qubit frequency
for $k_0d=\pi/2$ (dashed blue line in Fig.\ref{fig4}a). In this
case, the radiation propagates from left to right, from the first,
excited qubit, to the second, unexcited qubit, and does not
penetrate behind the second qubit. The second qubit acts as an
ideal mirror at this frequency. The same result was obtained in
\cite{Mak2003} by a different method.

\begin{figure}
  \includegraphics[width=8 cm]{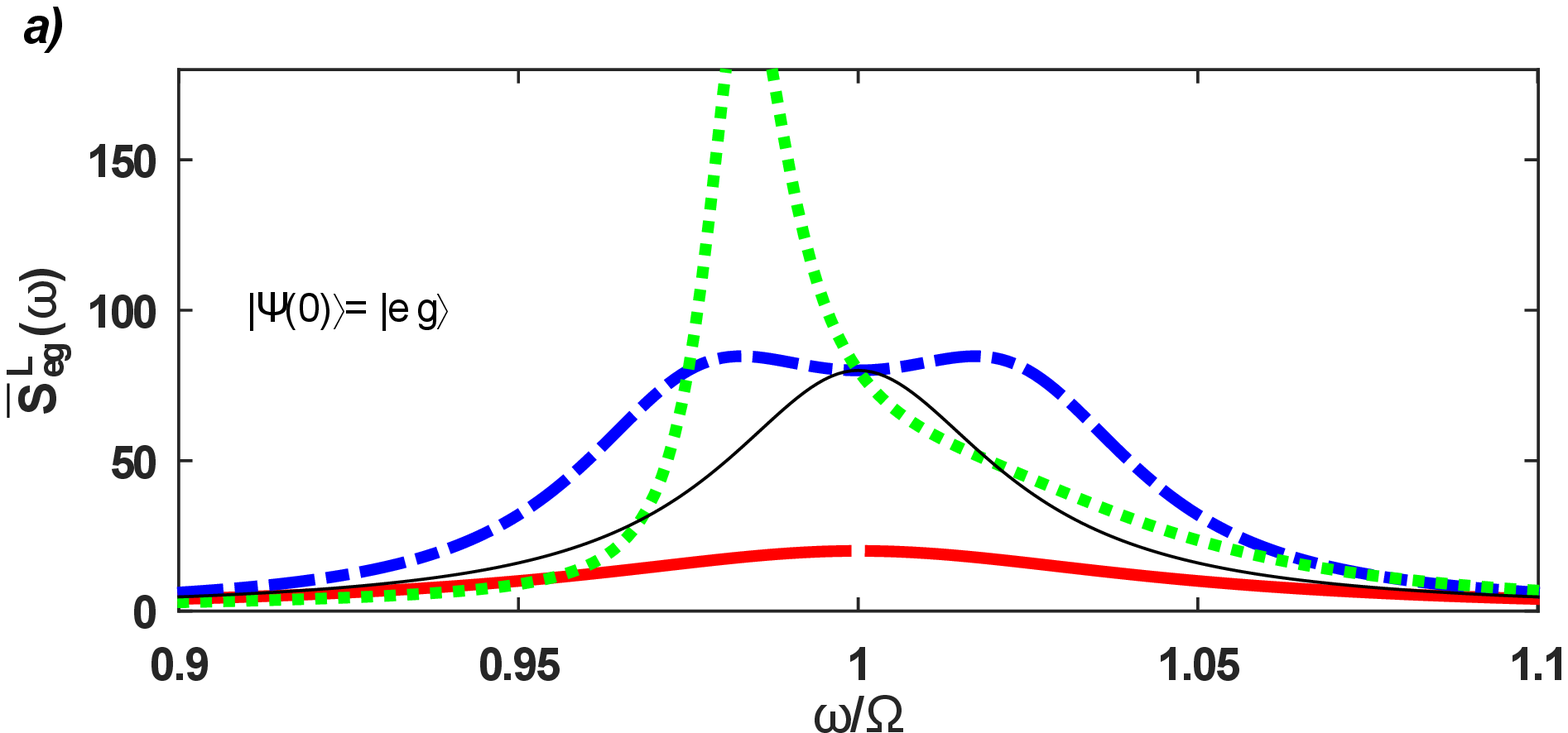}\\
  \includegraphics[width=8 cm]{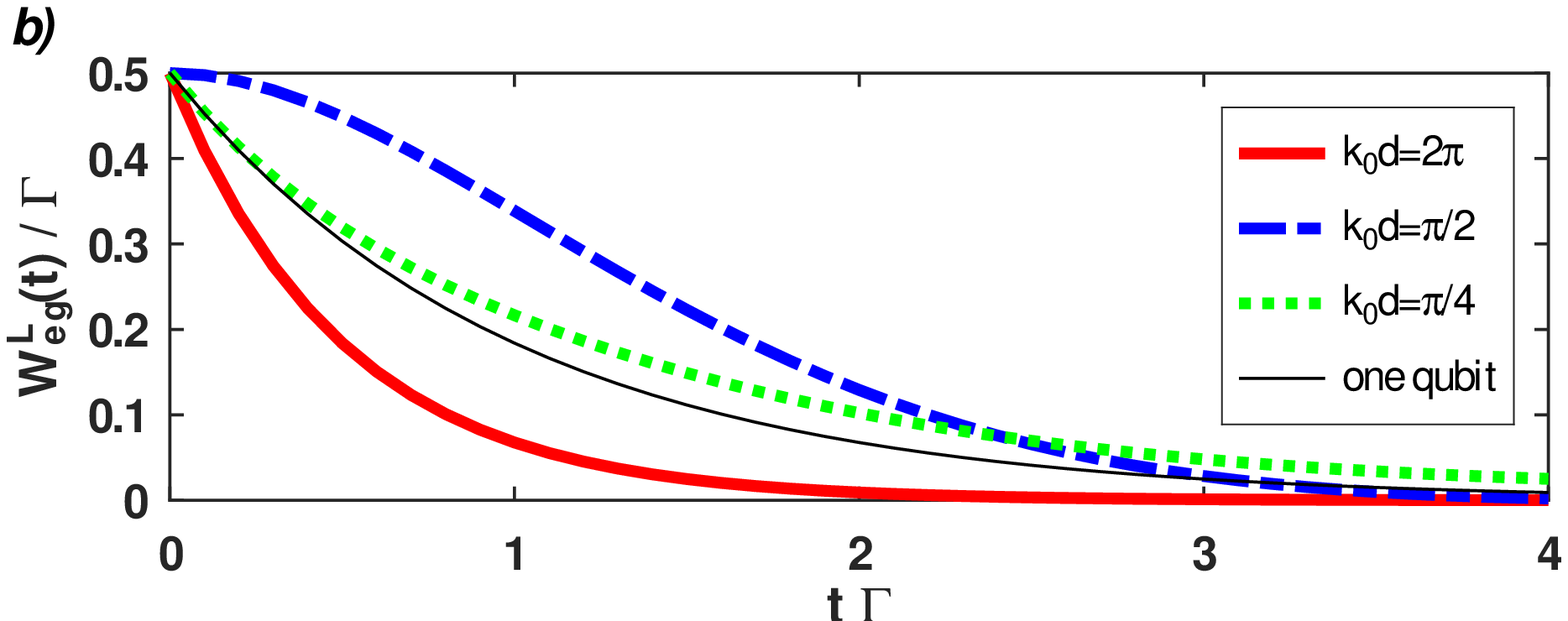}\\
  \caption{a) The backward radiation spectra $\overline{S}_{eg}^R(\omega)={S}^R_{eg}(\omega)2L\Omega/v_g$ for initially excited
  first qubit. b) The backward emission  decay rate, $W^R_{eg}/\Gamma$
  for initially excited first qubit; For comparison a single-qubit case
  is shown by black thin line. $\Gamma/\Omega=0.05$}\label{fig5}
\end{figure}

The radiation spectra and the emission rate for backward
scattering can be obtained from (\ref{6.32b}) and (\ref{6.32c})
for $k\approx -k_0$. The corresponding plots are shown in Fig.
\ref{fig5}a, Fig. \ref{fig5}b. The inter-peak distance (dashed
blue line in Fig. \ref{fig5}a, which is a measure of the photon
mediated coupling between qubits, is approximately $0.68\Gamma$.
Here, a superradiant decay also takes place for $k_0d=2\pi$, where
the interference terms in (\ref{6.32b}) and (\ref{6.32c}) are
equal to zero. In this case, backward radiation is the same as
that in the forward direction. However, for other values of $k_0d$
for which the interference terms are not zero, the backward
radiation is significantly different from the forward radiation
which is seen by the comparison between Fig. \ref{fig4}a, Fig.
\ref{fig4}b, and Fig. \ref{fig5}a, Fig. \ref{fig5}b.

If a second qubit is initially excited,
$\left|{\Psi(0)}\right\rangle =\left|{{g_1}{e_2}}\right\rangle $,
then the density matrix becomes ${\rho _S}(0) = \frac{1}{2}\left|
{S + A} \right\rangle \left\langle {S+A}\right|$. In this case,
the result is given by the equations (\ref{6.32a}), (\ref{6.32b}),
and (\ref{6.32c}) with the sign of last terms in these equations
being changed.  Therefore, for initially excited second qubit the
equations (\ref{6.32a}), (\ref{6.32b}), and (\ref{6.32c}) (as they
are written) describe the backward scattering for $k=+k_0$ and the
forward scattering for $k=-k_0$.

Therefore, from (\ref{6.32b}) we may conclude that, in general,
the probability to detect a photon by left or right detector is
not equal to each other since the backward and forward radiation
spectra are different. However, their sum,
$S_S(\omega)+S_A(\omega)$ is not direction sensitive since it does
not depend on the interference term.  The emission rate
(\ref{6.32c}), that is, the rate of the energy loss, is also
different for backward and forward emission, however their sum,
that is, the total emission rate, $W_S(t)+W_A(t)$ is not direction
sensitive.

\begin{widetext}
\subsection{Initial state with two excited qubits}

Here we consider the spectrum for the initial state with both
qubits being excited, $\left| {\Psi (0)} \right\rangle  = \left|
{e_1e_2} \right\rangle =\left|E\right\rangle$. The corresponding
density matrix is ${\rho _S}(0)= \left|E\right\rangle\left\langle
E \right|$. From (\ref{6.21}), (\ref{6.21b}) we obtain the
following result for the radiation spectrum:

\begin{equation}\label{6.37a}
\begin{gathered}
{\left\langle {a_k^\dag (t){a_k}(t)} \right\rangle _E} = \frac{{{\Gamma _ + }}}{{{\Gamma _ - }}}{\left\langle {a_k^\dag (t){a_k}(t)} \right\rangle _S} + \frac{{{\Gamma _ - }}}{{{\Gamma _ + }}}{\left\langle {a_k^\dag (t){a_k}(t)} \right\rangle _A}\\
 + \frac{{{\upsilon _g}}}{{2L}}\frac{{\Gamma _ + ^2 + \Gamma _ - ^2}}{{2\Gamma }}\left( {{e^{ - 2\Gamma t}} - 1} \right)\left[ {\frac{{{e^{ - i{k_0}d}}}}{{{\Gamma _ + }\left( {1 + i\sin ({k_0}d)} \right)\left( {i{\delta _ - } - {\Gamma _ - }/2} \right)}}} \right. - \frac{{{e^{ - i{k_0}d}}}}{{{\Gamma _ - }\left( {1 - i\sin ({k_0}d)} \right)\left( {i{\delta _ + } - {\Gamma _ + }/2} \right)}}\\
 + \left. {\frac{{{e^{i{k_0}d}}}}{{{\Gamma _ - }\left( {1 + i\sin ({k_0}d)} \right)\left( {i{\delta _ + } - {\Gamma _ - }/2 - \Gamma } \right)}} - \frac{{{e^{i{k_0}d}}}}{{{\Gamma _ + }\left( {1 - i\sin ({k_0}d)} \right)\left( {i{\delta _ - } - {\Gamma _ + }/2 - \Gamma } \right)}}} \right]\\
 + \frac{{{\upsilon _g}}}{{2L\;{\Gamma _ + }}}\frac{{{e^{i{k_0}d}}}}{{\left( {1 - i\sin ({k_0}d)} \right)}}\frac{{\left( {{e^{ - \left( {i{\delta _ - } + {{{\Gamma _ - }} \mathord{\left/
 {\vphantom {{{\Gamma _ - }} 2}} \right.
 \kern-\nulldelimiterspace} 2}} \right)t}} - 1} \right)\left( {\Gamma _ - ^2 + \Gamma _ + ^2{e^{\left( {i{\delta _ - } - {\Gamma _ + }/2 - \Gamma } \right)t}}} \right)}}{{\left( {i{\delta _ - } + {\Gamma _ - }/2} \right)\left( {i{\delta _ - } - {\Gamma _ + }/2 - \Gamma } \right)}}\\
 - \frac{{{\upsilon _g}}}{{2L\;{\Gamma _ - }}}\frac{{{e^{i{k_0}d}}}}{{\left( {1 + i\sin ({k_0}d)} \right)}}\frac{{\left( {{e^{ - \left( {i{\delta _ + } + {\Gamma _ + }/2} \right)t}} - 1} \right)\left( {\Gamma _ + ^2 + \Gamma _ - ^2{e^{\left( {i{\delta _ + } - {\Gamma _ - }/2 - \Gamma } \right)t}}} \right)}}{{\left( {i{\delta _ + } + {\Gamma _ + }/2} \right)\left( {i{\delta _ + } - {\Gamma _ - }/2 - \Gamma } \right)}}\\
 + \frac{{{\upsilon _g}}}{{2L\;{\Gamma _ - }}}\frac{{{e^{ - i{k_0}d}}}}{{\left( {1 - i\sin ({k_0}d)} \right)}}\frac{{\left( {{e^{ - \left( {i{\delta _ + } + {\Gamma _ - }/2 + \Gamma } \right)t}} - 1} \right)\left( {\Gamma _ - ^2 + \Gamma _ + ^2{e^{\left( {i{\delta _ + } - {\Gamma _ + }/2} \right)t}}} \right)}}{{\left( {i{\delta _ + } - {\Gamma _ + }/2} \right)\left( {i{\delta _ + } + {\Gamma _ - }/2 + \Gamma } \right)}}\\
 - \frac{{{\upsilon _g}}}{{2L\;{\Gamma _ + }}}\frac{{{e^{ - i{k_0}d}}}}{{\left( {1 + i\sin ({k_0}d)} \right)}}\frac{{\left( {{e^{ - \left( {i{\delta _ - } + {\Gamma _ + }/2 + \Gamma } \right)t}} - 1} \right)\left( {\Gamma _ + ^2 + \Gamma _ - ^2{e^{\left( {i{\delta _ - } - {\Gamma _ - }/2} \right)t}}} \right)}}{{\left( {i{\delta _ - } - {\Gamma _ - }/2} \right)\left( {i{\delta _ - } + {\Gamma _ + }/2 + \Gamma } \right)}}
\end{gathered}
\end{equation}
By taking time in (\ref{6.37a}) to infinity, we get the radiation
spectral density:
\begin{equation}\label{6.37b}
\begin{gathered}
{S_E}(\omega ) = \frac{{{\Gamma _ + }}}{{{\Gamma _ - }}}{S_S}(\omega ) + \frac{{{\Gamma _ - }}}{{{\Gamma _ + }}}{S_A}(\omega ) + \frac{{{\upsilon _g}}}{{2L}}\frac{{\Gamma _ + ^2 + \Gamma _ - ^2}}{{2\Gamma \left( {1 + {{\sin }^2}({k_0}d)} \right)}}\left[ {\frac{{{e^{i{k_0}d}}\left( {1 + i\sin ({k_0}d)} \right)}}{{{\Gamma _ + }\left( {i{\delta _ - } - {\Gamma _ + }/2 - \Gamma } \right)}} - \frac{{{e^{ - i{k_0}d}}\left( {1 - i\sin ({k_0}d)} \right)}}{{{\Gamma _ + }\left( {i{\delta _ - } - {\Gamma _ - }/2} \right)}}} \right.\\
 + \left. {\frac{{{e^{ - i{k_0}d}}\left( {1 + i\sin ({k_0}d)} \right)}}{{{\Gamma _ - }\left( {i{\delta _ + } - {\Gamma _ + }/2} \right)}} - \frac{{{e^{i{k_0}d}}\left( {1 - i\sin ({k_0}d)} \right)}}{{{\Gamma _ - }\left( {i{\delta _ + } - {\Gamma _ - }/2 - \Gamma } \right)}}} \right]\\
 + \frac{{{\upsilon _g}}}{{2L}}\frac{1}{{\left( {1 + {{\sin }^2}({k_0}d)} \right)}}\left[ {\frac{{\Gamma _ + ^2}}{{{\Gamma _ + }}}\frac{{{e^{ - i{k_0}d}}\left( {1 - i\sin ({k_0}d)} \right)}}{{\left( {i{\delta _ - } - {\Gamma _ - }/2} \right)\left( {i{\delta _ - } + {\Gamma _ + }/2 + \Gamma } \right)}} - \frac{{\Gamma _ - ^2}}{{{\Gamma _ + }}}\frac{{{e^{i{k_0}d}}\left( {1 + i\sin ({k_0}d)} \right)}}{{\left( {i{\delta _ - } + {\Gamma _ - }/2} \right)\left( {i{\delta _ - } - {\Gamma _ + }/2 - \Gamma } \right)}}} \right]\\
 + \frac{{{\upsilon _g}}}{{2L}}\frac{1}{{\left( {1 + {{\sin }^2}({k_0}d)} \right)}}\left[ {\frac{{\Gamma _ + ^2}}{{{\Gamma _ - }}}\frac{{{e^{i{k_0}d}}\left( {1 - i\sin ({k_0}d)} \right)}}{{\left( {i{\delta _ + } + {\Gamma _ + }/2} \right)\left( {i{\delta _ + } - {\Gamma _ - }/2 - \Gamma } \right)}} - \frac{{\Gamma _ - ^2}}{{{\Gamma _ - }}}\frac{{{e^{ - i{k_0}d}}\left( {1 + i\sin ({k_0}d)} \right)}}{{\left( {i{\delta _ + } - {\Gamma _ + }/2} \right)\left( {i{\delta _ + } + {\Gamma _ - }/2 + \Gamma } \right)}}} \right]
\end{gathered}
\end{equation}
For the emission rate we obtain from (\ref{6.37a}):
\begin{equation}\label{6.37c}
{W_E}(t) = \frac{1}{2}\frac{{\Gamma _ + ^2}}{{{\Gamma _ - }}}
{e^{ - {\Gamma _ + }t}} + \frac{1}{2}\frac{{\Gamma _ - ^2}}
{{{\Gamma _ + }}}{e^{ - {\Gamma _ - }t}} - \frac{{4\Gamma {{\cos }^2}({k_0}d)}}
{{1 - {{\cos }^2}({k_0}d)}}{e^{ - 2\Gamma t}}
\end{equation}

As might appear at the first sight the expressions
(\ref{6.37a}-\ref{6.37c}) may take the infinite values for $k_0d =
n\pi$ due to the widths $\Gamma_-,\Gamma_+$ in the denominator.
However, a close inspection of these equations reveals that at
these points the numerator is also zero.  As previously, we can
obtain the right solution if we put $k_0d = n\pi$ directly in the
equations (\ref{6.6}) and (\ref{6.7}), or by expanding $cos(k_0d)$
near $k_0d = n\pi$ in (\ref{6.37a}-\ref{6.37c}). For example, for
$k_0d = 2\pi$ we find:

\begin{equation}\label{6.38a}
    \begin{gathered}
{\left. {{{\left\langle {a_k^\dag (t){a_k}(t)} \right\rangle }_E}} \right|_{{k_0}d = 2\pi }} = 4\frac{{{\upsilon _g}\Gamma }}{{2L}}\frac{{\left( {{e^{ - \left( {i\delta  + \Gamma } \right)t}} - 1} \right)\left( {{e^{\left( {i\delta  - \Gamma } \right)t}} - 1} \right)}}{{{\delta ^2} + {\Gamma ^2}}} + 2\frac{{{\upsilon _g}\Gamma }}{{2L}}\frac{{{\Gamma ^2}}}{{i\delta {{\left( {i\delta  - \Gamma } \right)}^2}\left( {i\delta  - 2\Gamma } \right)}}\\
 + 2\frac{{{\upsilon _g}\Gamma }}{{2L}}\frac{{{e^{ - 2\Gamma t}}}}{{{{\left( {i\delta  - \Gamma } \right)}^2}}} - 2\frac{{{\upsilon _g}\Gamma }}{{2L}}\frac{{{e^{ - \left( {i\delta  + 2\Gamma } \right)t}} - 1}}{{i\delta \left( {i\delta  + 2\Gamma } \right)}} - 2\frac{{{\upsilon _g}\Gamma }}{{2L}}\frac{{{e^{\left( {i\delta  - 2\Gamma } \right)t}}}}{{i\delta \left( {i\delta  - 2\Gamma } \right)}}\\
 - 2\frac{{{\upsilon _g}\Gamma }}{{2L}}\frac{{2\Gamma }}{{{{\left( {i\delta  - \Gamma } \right)}^2}}}\frac{{{e^{ - \left( {i\delta  + \Gamma } \right)t}} - 1}}{{\left( {i\delta  + \Gamma } \right)}} + 2\frac{{{\upsilon _g}\Gamma }}{{2L}}\frac{{2\Gamma }}{{\left( {i\delta  - \Gamma } \right)}}\frac{{{e^{\left( {i\delta  - \Gamma } \right)t}} - {e^{ - 2\Gamma t}}}}{{{{\left( {i\delta  + \Gamma } \right)}^2}}} + 2\frac{{{\upsilon _g}\Gamma }}{{2L}}\frac{{{e^{ - 2\Gamma t}}}}{{{\delta ^2} + {\Gamma ^2}}}2\Gamma t
\end{gathered}
\end{equation}

\end{widetext}

\begin{equation}\label{6.38b}
   {\left. {{S_E}(\omega )} \right|_{{k_0}d
= 2\pi }} =  \frac{{{\upsilon _g}}}{{2L}}\frac{{6\Gamma \left(
{{\delta ^2}
  + 2{\Gamma ^2}} \right)}}{{\left( {{\delta ^2}
  + {\Gamma ^2}} \right)\left( {{\delta ^2} + 4{\Gamma ^2}} \right)}}
\end{equation}

\begin{equation}\label{6.38c}
    {\left. {{W_E}(t)} \right|_{{k_0}d = 2\pi }}
 = \left( {1 + 2\Gamma t} \right)\Gamma {e^{ - 2\Gamma t}}
\end{equation}

\begin{figure}
  \includegraphics[width=8 cm]{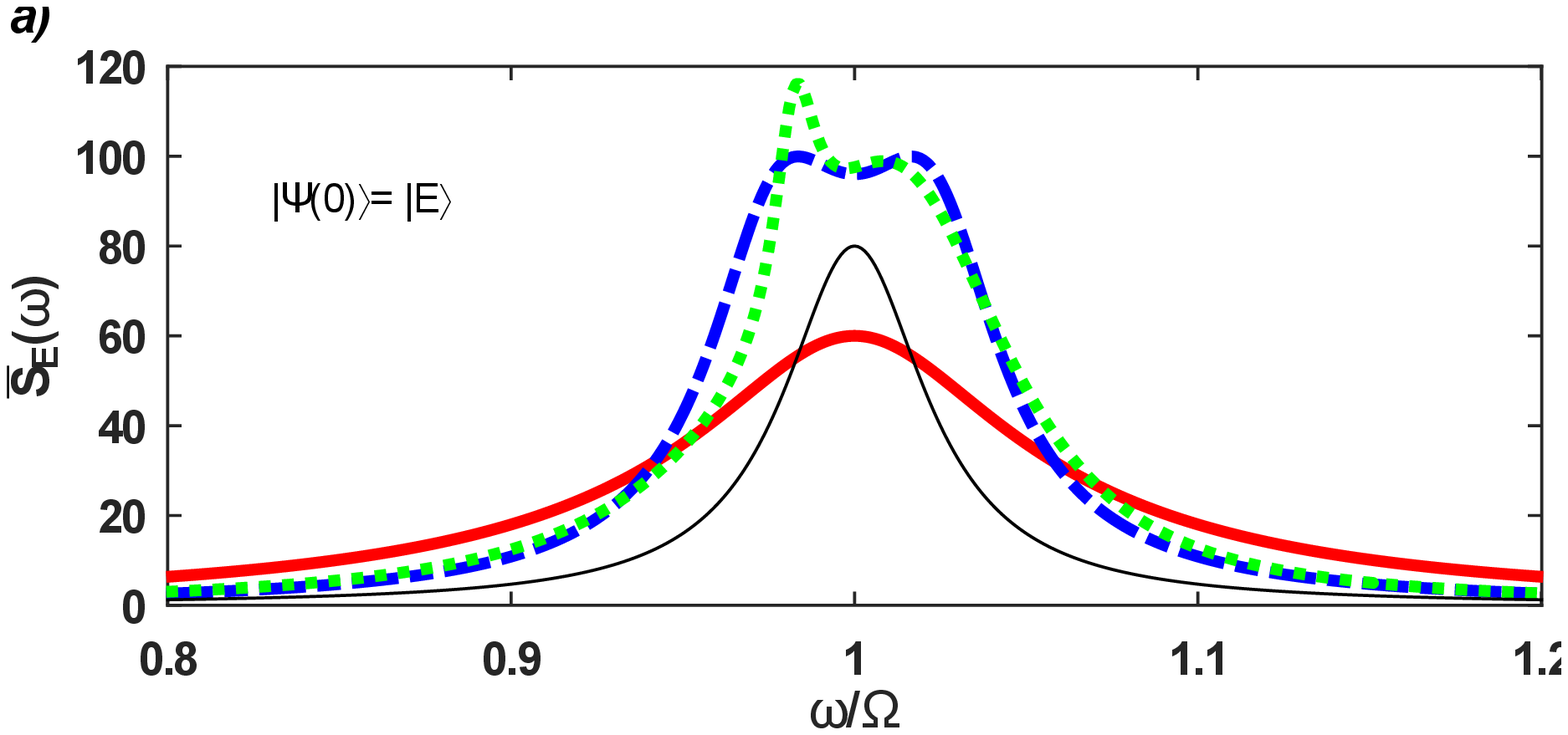}\\
  \includegraphics[width=8 cm]{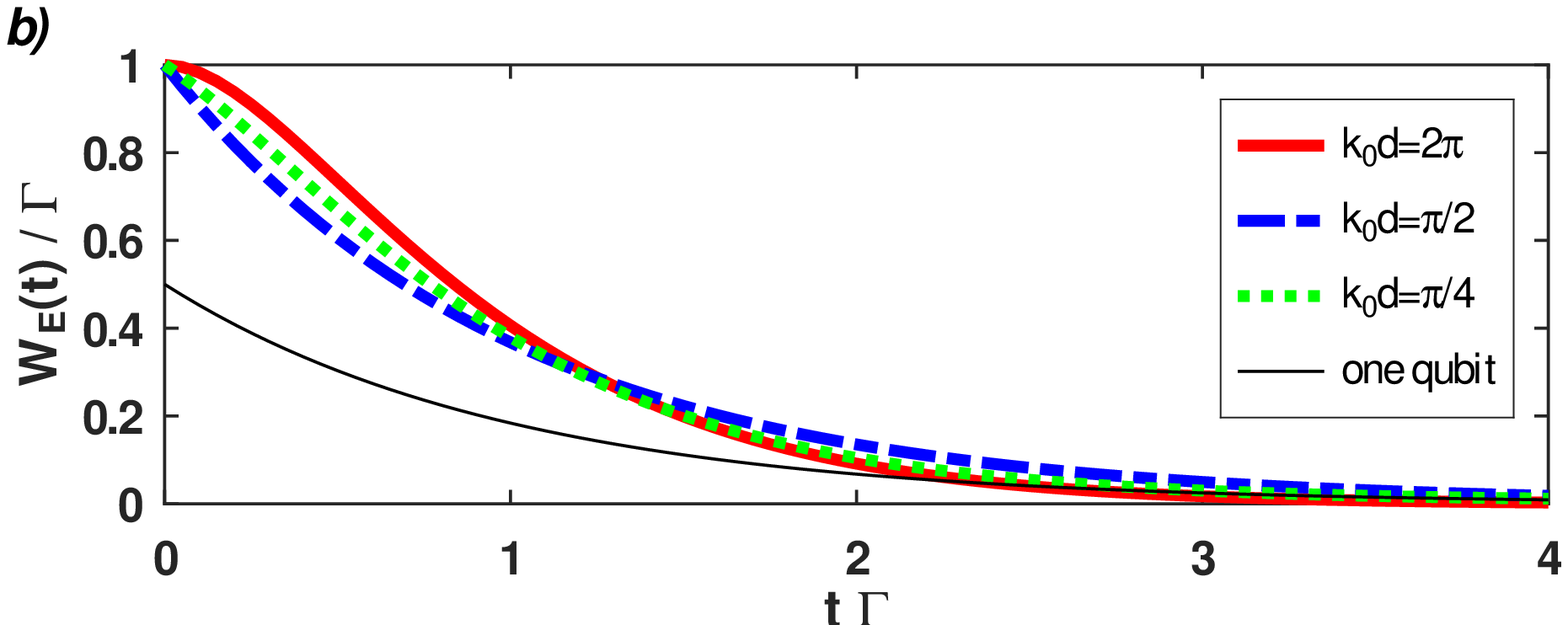}\\
  \caption{a) Radiation spectra $\overline{S}_{E}(\omega)={S}_{E}(\omega)2L\Omega/v_g$, expression (\ref{6.37b}), for  two initially excited qubits (\ref{6.37b}) for different effective distances; b) Emission decay rate, $W_E(t)/\Gamma$ of two excited qubits (\ref{6.37c}); $\Gamma/\Omega=0.05$.} \label{fig6}
\end{figure}

For this case, the radiation spectrum for several values of $k_0d$
is shown in Fig.\ref{fig6}a. For $k_0d = n\pi$, where $n$ is any
integer, we obtain a single-peak Lorentzian line, though the
analytical function (\ref{6.38b}) is more complex. For
$k_0d=\pi/2$ there are two peaks with a small separation at the
top. Here, the inter-peak distance is approximately $0.66\Gamma$.
For $k_0d=\pi/4$ there are two asymmetrical peaks which can be a
signature of Fano resonance in the system.

The plots of emission decay rate for these values of $k_0d$ are
shown in Fig.\ref{fig6}b. It is seen that the emission rates for
two initially excited qubits are noticeably faster than the decay
rate of a single qubit. Obviously, this is a signature of
superradiant emission.


\subsection{Initial states with qubits superposition}
\subsubsection {First qubit is in a
superposition state, a second qubit is in a ground state}

\begin{multline}\label{6.39a}
\left| {\Psi (0)} \right\rangle  = \left| {{s_1}} \right\rangle  \otimes \left| {{g_2}} \right\rangle  = \frac{1}{{\sqrt 2 }}\left( {\left| {{e_1}} \right\rangle  + \left| {{g_1}} \right\rangle } \right) \otimes \left| {{g_2}} \right\rangle \\
 = \frac{1}{2}\left| S \right\rangle  - \frac{1}{2}\left| A \right\rangle  + \frac{1}{{\sqrt 2 }}\left| G \right\rangle
\end{multline}
The corresponding initial density matrix is given by:
\begin{multline}\label{6.39b}
{\rho _S}(0) = \frac{1}{4}\left( {\left| S \right\rangle \left\langle S \right| + \left| A \right\rangle \left\langle A \right| - \left| S \right\rangle \left\langle A \right| - \left| A \right\rangle \left\langle S \right|} \right)\\
 + \frac{1}{{2\sqrt 2 }}\left( {\left| S \right\rangle \left\langle G \right| - \left| A \right\rangle \left\langle G \right| + \left| G \right\rangle \left\langle S \right| - \left| G \right\rangle \left\langle A \right|} \right) + \frac{1}{2}\left| G \right\rangle \left\langle G \right|
\end{multline}
The first line in (\ref{6.39b}) is a half of the density matrix of
the state $\left|{eg} \right\rangle$ (\ref{6.30}), therefore,  we
get the same spectrum and other related parameters similar to
those for the first excited qubit (\ref{6.32a}-\ref{6.32c}), but
reduced by the factor of two:
\begin{equation}\label{6.40}
\begin{gathered}
{\left\langle {a_k^\dag (t){a_k}(t)} \right\rangle _{{s_1}{g_2}}} = \frac{1}{2}{\left\langle {a_k^\dag (t){a_k}(t)} \right\rangle _{eg}};\\
{S_{{s_1}{g_2}}}(\omega ) = \frac{1}{2}{S_{eg}}(\omega );\quad \quad {W_{{s_1}{g_2}}}(t) = \frac{1}{2}{W_{eg}}(t);
\end{gathered}
\end{equation}
The second line in (\ref{6.39b}) which describes the transitions
to the ground state does not contribute to $\left\langle {a_k^\dag
(t){a_k}(t)} \right\rangle$ (expressions (\ref{6.21}),
(\ref{6.21b})).

Therefore the state with the first qubit prepared in a
superposition state and the second one in a ground state shows the
spectral properties identical to those shown in Figs. \ref{fig4},
\ref{fig5}, but on a smaller scale.


\subsubsection{First qubit is in a
superposition state, a second qubit is in an excited state}

\begin{multline}\label{6.42a}
\left| {\Psi (0)} \right\rangle  = \left| {{s_1}} \right\rangle  \otimes \left| {{e_2}} \right\rangle  = \frac{1}{{\sqrt 2 }}\left( {\left| {{e_1}} \right\rangle  + \left| {{g_1}} \right\rangle } \right) \otimes \left| {{e_2}} \right\rangle \\
 = \frac{1}{{\sqrt 2 }}\left| E \right\rangle  + \frac{1}{2}\left| S \right\rangle  + \frac{1}{2}\left| A \right\rangle
\end{multline}
and the corresponding initial density matrix is:
\begin{multline}\label{6.42b}
{\rho _S}(0) = \frac{{\left| E \right\rangle \left\langle E \right|}}{2} + \frac{{\left| S \right\rangle \left\langle S \right| + \left| A \right\rangle \left\langle A \right| + \left| S \right\rangle \left\langle A \right| + \left| A \right\rangle \left\langle S \right|}}{4}\\
 + \frac{{\left| E \right\rangle \left\langle S \right| + \left| E \right\rangle \left\langle A \right| + \left| S \right\rangle \left\langle E \right| + \left| A \right\rangle \left\langle E \right|}}{{2\sqrt 2 }}
\end{multline}
As in the previous example, the first line in (\ref{6.42b})
corresponds to the initial states already considered above.
Therefore, we can construct the spectrum and emission rate using
(\ref{6.37a}) and $ge$ counterpart of (\ref{6.32a}) (see the last
paragraph in Sec.VIIB):
\begin{equation}\label{6.43}
\begin{gathered}
{\left\langle {a_k^\dag (t){a_k}(t)} \right\rangle _{{s_1}{e_2}}}
 = \frac{1}{2}{\left\langle {a_k^\dag {a_k}} \right\rangle _E}
  + \frac{1}{2}{\left\langle {a_k^\dag {a_k}} \right\rangle _{ge}},\quad \quad \\
{S_{{s_1}{e_2}}}(\omega ) = \frac{1}{2}{S_E}(\omega ) + \frac{1}{2}{S_{ge}}(\omega ),\\
{W_{{s_1}{e_2}}}(t) = \frac{1}{2}{W_E}(t) + \frac{1}{2}{W_{ge}}(t),
\end{gathered}
\end{equation}
where  ${\left\langle {a_k^\dag {a_k}} \right\rangle _{ge}}$,
${S_{ge}}(\omega )$, and ${W_{ge}}(t)$ are given by the equations
(\ref{6.32a}), (\ref{6.32b}), and (\ref{6.32c}) with the sign of
the interference term in these equations being changed.

\begin{figure}
  \includegraphics[width=8 cm]{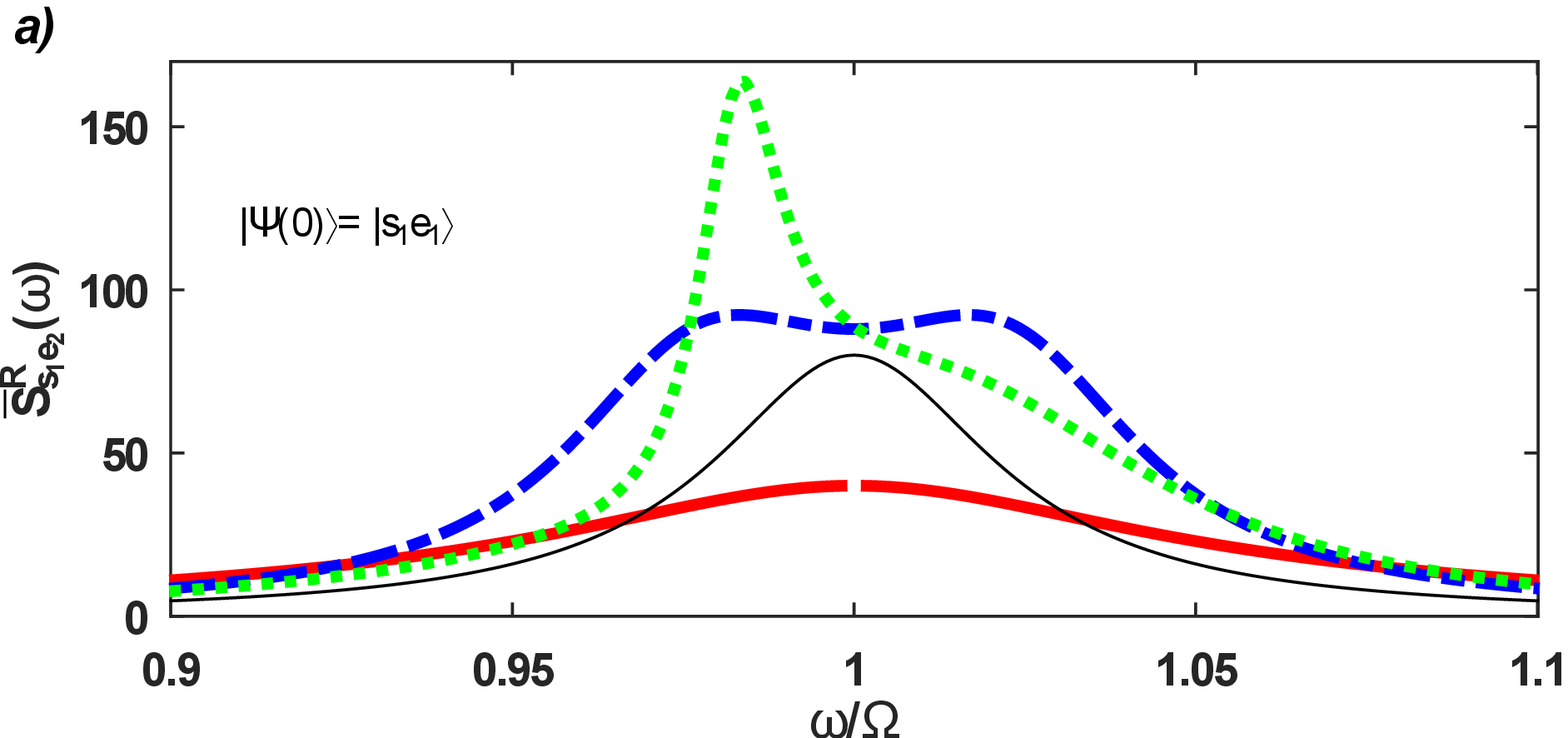}\\
  \includegraphics[width=8 cm]{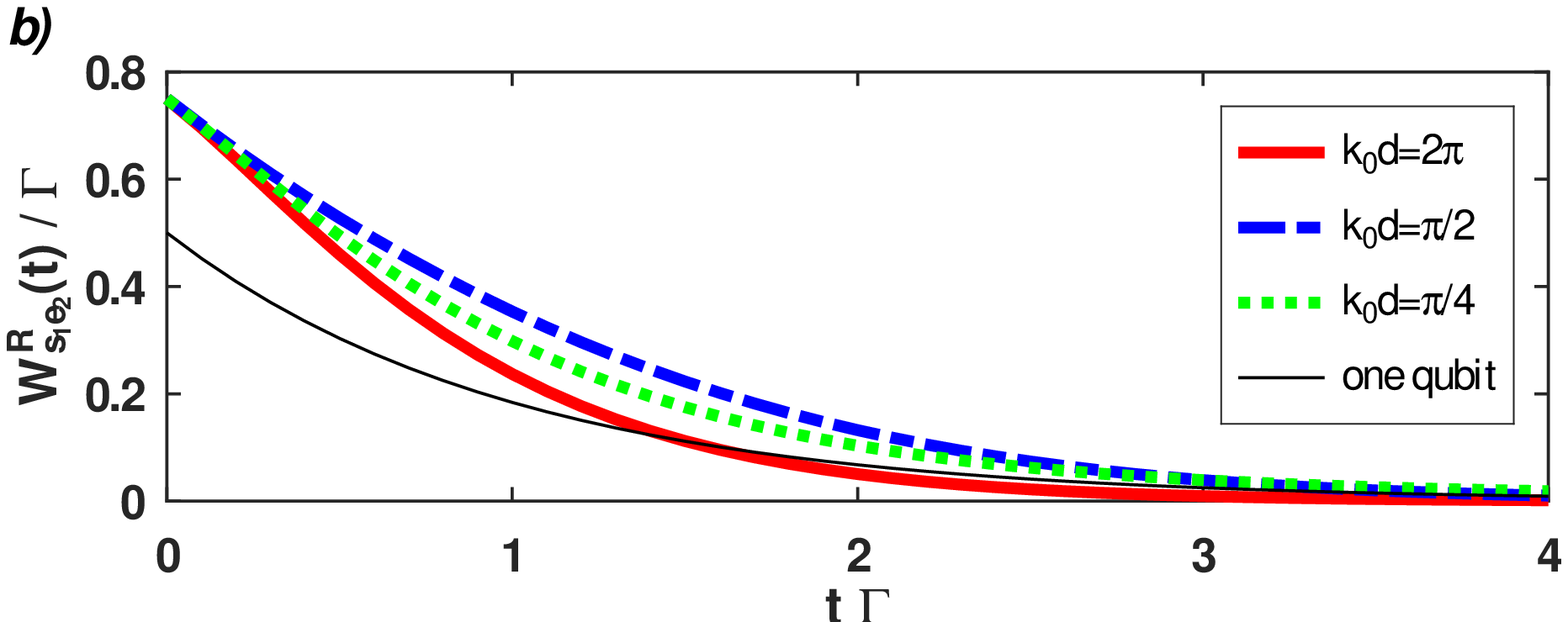}\\
\caption{a) Forward radiation spectra
$\overline{S^L}_{s_1,e_2}(\omega)={S^L}_{s_1,e_2}(\omega)2L\Omega/v_g$
for the initial state with the first qubit being in a
superposition state and the second one being in the excited state
for the different $k_0d$; b) The forward emission decay rates $W^L_{s_1,e_2}/\Gamma$ for the same initial state; The single-qubit decay rate is $\Gamma/\Omega=0.05$.} \label{fig7}
\end{figure}

\begin{figure}
  \includegraphics[width=8 cm]{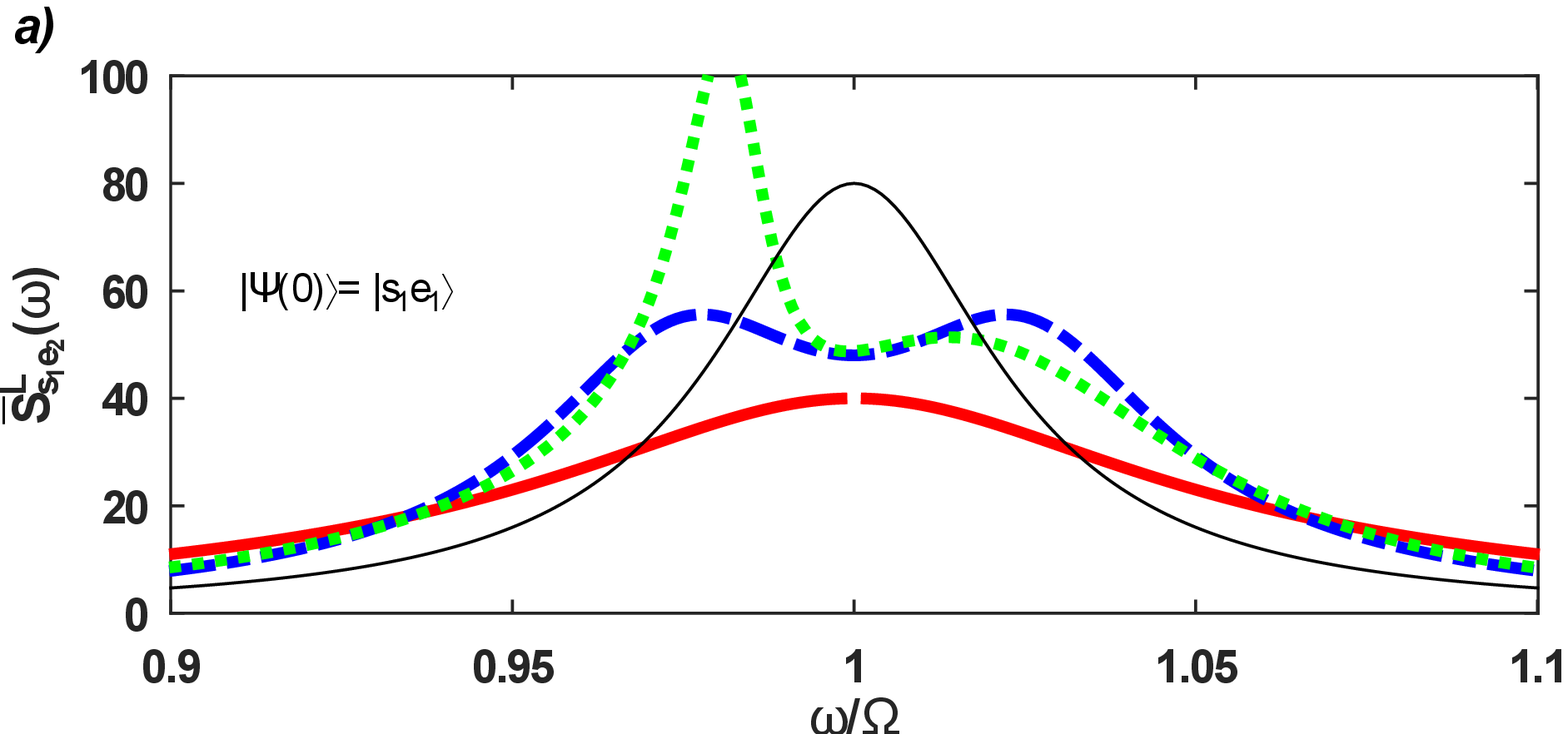}\\
  \includegraphics[width=8 cm]{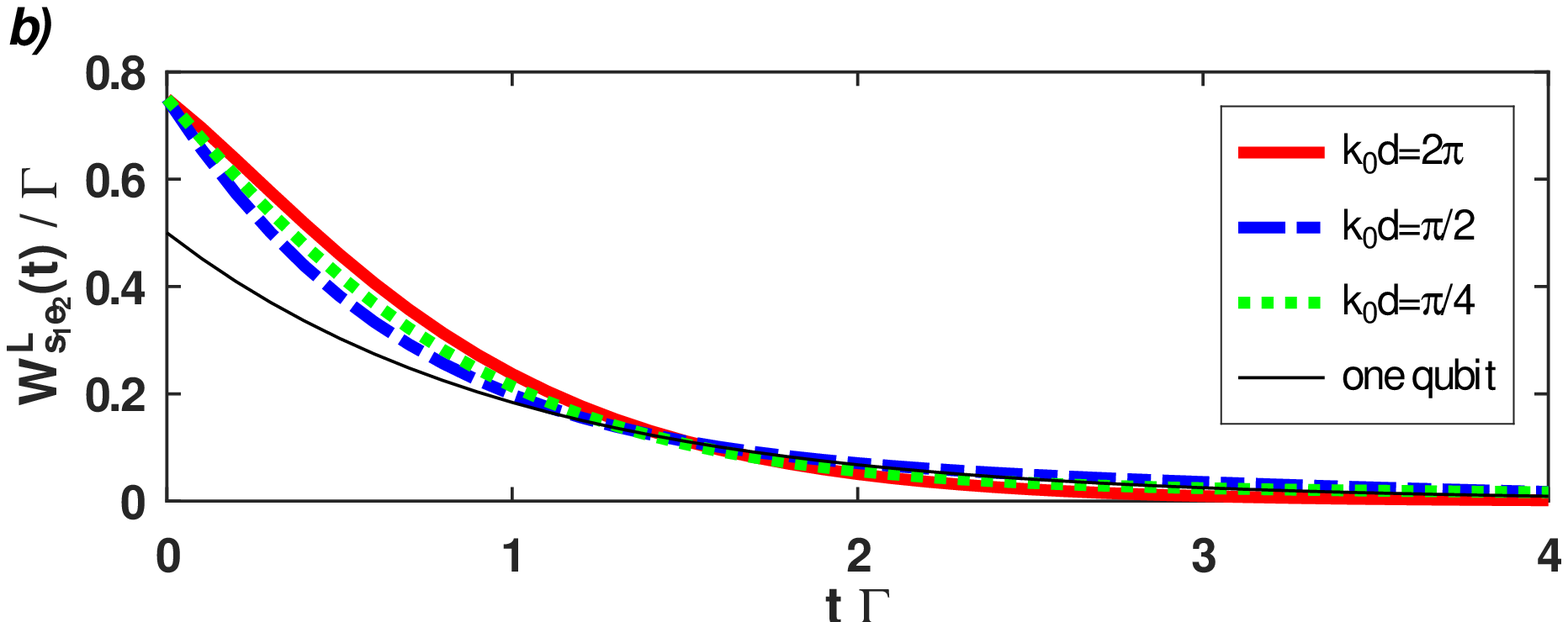}\\
\caption{a) Backward radiation spectra
$\overline{S^R}_{s_1,e_2}(\omega)={S^R}_{s_1,e_2}(\omega)2L\Omega/v_g$
for the initial state with the first qubit being in a
superposition state and the second one being in the excited state
for the different $k_0d$; b) Backward emission decay rates $W^R_{s_1,e_2}/\Gamma$ for the same initial state; The single-qubit decay rate is $\Gamma/\Omega=0.05$} \label{fig8}
\end{figure}

Thus, the spectrum for the initial state (\ref{6.42a}) is a
combination of the spectrum of two-excited qubits and that of a
first excited qubit. Here, the probabilities for the photon to be
emitted in left and right directions are different. The forward
and backward radiation spectra and emission rates for this case
are presented in Figs. \ref{fig7}a, \ref{fig8}a, and Figs.
\ref{fig7}b, \ref{fig8}b, respectively.

\subsubsection{Both qubits are initially prepared
in a superposition state}

\begin{multline}\label{6.45a}
\left| {\Psi (0)} \right\rangle  = \frac{1}{{\sqrt 2 }}\left( {\left| {{e_1}}
 \right\rangle  + \left| {{g_1}} \right\rangle } \right)
 \otimes \frac{1}{{\sqrt 2 }}\left( {\left| {{e_2}} \right\rangle
  + \left| {{g_2}} \right\rangle } \right)\\
 = \frac{1}{2}\left| E \right\rangle  + \frac{1}{{\sqrt 2 }}\left| S
 \right\rangle  + \frac{1}{2}\left| G \right\rangle
\end{multline}

with the initial density matrix:

\begin{multline}\label{6.45b}
{\rho _S}(0) = \frac{\left| E \right\rangle \left\langle
E\right|}{4}
 + \frac{{\left| S \right\rangle \left\langle S \right|}}{2}\\
 +\frac{ \left| E \right\rangle \left\langle G \right|
  + \left| G \right\rangle \left\langle E \right|
   + \left| G \right\rangle \left\langle G \right|}{4}\\
 + \frac{1}{{2\sqrt 2 }}\left( {\left| E \right\rangle \left\langle S \right|
  + \left| S \right\rangle \left\langle E \right|
   + \left| S \right\rangle \left\langle G \right|
    + \left| G \right\rangle \left\langle S \right|} \right)
\end{multline}

\begin{figure}
  \includegraphics[width=8 cm]{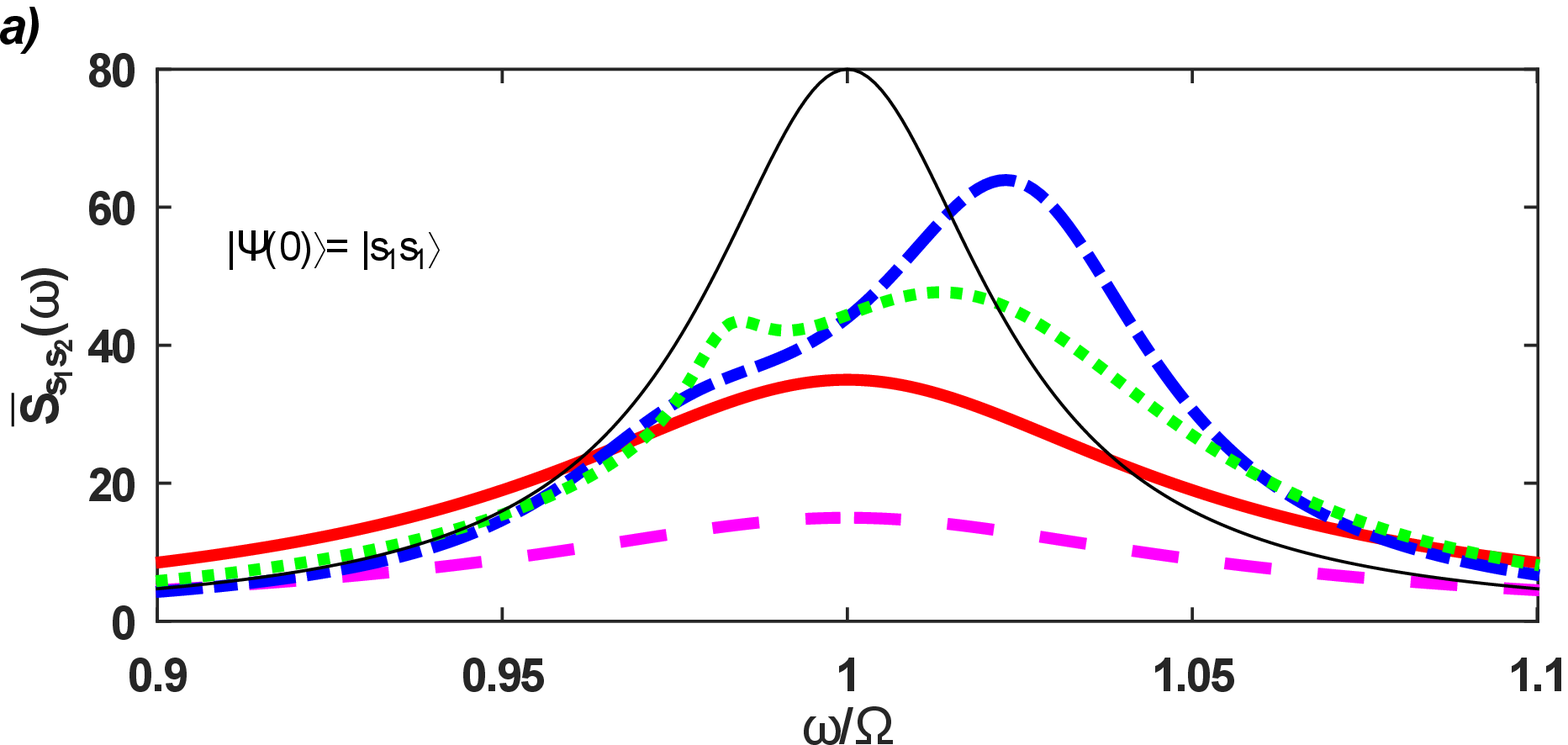}\\
   \includegraphics[width=8 cm]{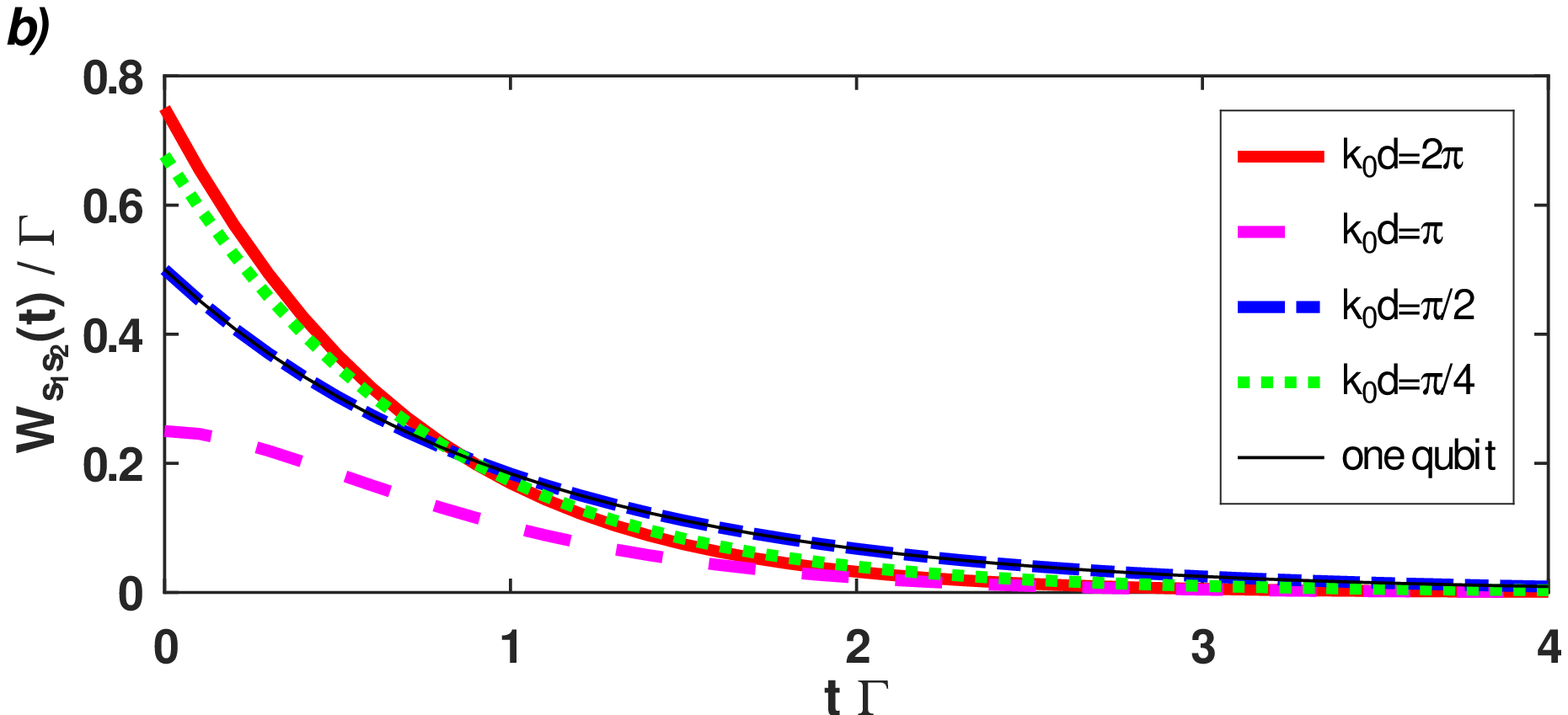}\\
  \caption{a) Radiation spectra $\overline{S}_{s_1,s_2}(\omega)={S}_{s_1,s_2}(\omega)2L\Omega/v_g$ for the state in which every qubit is initially in a superposition state;
b) The emission decay rate $W_{s_1,s_2}/\Gamma$ for the initial superposition state; $\Gamma/\Omega=0.05$.}\label{fig9}
\end{figure}

As it follows from (\ref{6.21}), (\ref{6.21b}), only the first
line in (\ref{6.45b}) contributes to the radiation spectrum, which
can be presented as a combination of a two-excited qubit state
(\ref{6.37a}-\ref{6.37c}) and a symmetrical state (\ref{6.27a},
\ref{6.27b}, \ref{6.27c}):
\begin{equation}\label{6.46}
\begin{gathered}
{\left\langle {a_k^\dag (t){a_k}(t)} \right\rangle _{{s_1}{s_2}}} = \frac{1}{4}{\left\langle {a_k^\dag {a_k}} \right\rangle _E} + \frac{1}{2}{\left\langle {a_k^\dag {a_k}} \right\rangle _S},\quad \quad \\
{S_{{s_1}{s_2}}}(\omega ) = \frac{1}{4}{S_E}(\omega ) + \frac{1}{2}{S_S}(\omega ),\\
{W_{{s_1}{s_2}}}(t) = \frac{1}{4}{W_E}(t) + \frac{1}{2}{W_S}(t).
\end{gathered}
\end{equation}
For this case, the probabilities to find the photon in left or
right detectors are the same. The characteristic plots for this
case are presented in Figs. \ref{fig9}a, b. We see from
Fig. \ref{fig9}b that the plot for $k_0d=\pi/2$ is superimposed on
a single qubit plot. It means that the areas of corresponding
spectral lines (Fig. \ref{fig9}a) are equal to each other,
although their line shapes are different. Another feature is the
existence of a subradiant state for $k_0d=\pi$, dashed purple line
in Figs. \ref{fig9}a, \ref{fig9}b.

As a concluding remarks to this subsection we note that as can be
seen from Fig. \ref{fig7}a, the radiation spectra for the initial
state with the first qubit being in a superposition state and the
other being in an excited state are very similar to the spectrum
of two excited qubits shown in Fig. \ref{fig6}a. On the other
hand, when both qubits are prepared in a superposition state, the
spectrum changes significantly, and only for $k_0d = n\pi$
similarity is retained. Note that the line width of both spectrum
(\ref{6.43}) and (\ref{6.46}) for $k_0d = 2\pi$ is identical.
Moreover, it matches with the line width of the spectrum of two
excited qubits (\ref{6.38b}) (red line in Fig. \ref{fig6}a).

\section{Conclusion}
In this paper we investigate superradiant and subradiant
properties of the photon emission spectra for a two-qubit system
coupled to one dimensional open waveguide. We obtain the general
expression which allows us to calculate the radiation spectra for
arbitrary initial configuration of a two-qubit system. We obtain
the explicit expressions for the photon radiation spectra and the
emission decay rates for different initial two-qubit
configurations with one and two excitations. We show that the line
shape of the photon radiation spectra and the emission decay rate,
that is, the rate of the energy loss depend significantly on the
effective distance between qubits, $k_0d$.

We believe that the results obtained in this paper may have
practical applications in quantum information technologies
including a control and optimization of the two-qubit entangling
gates necessary for the realization of arbitrary unitary
operations needed for quantum computation.

\begin{acknowledgments}
The work is supported by the Ministry of Science and Higher
Education of Russian Federation under the project FSUN-2020-0004
and by the Foundation for the Advancement of Theoretical Physics
and Mathematics "BASIS".
\end{acknowledgments}

\end{document}